\documentclass{jfm}
\usepackage{graphicx}
\usepackage{epstopdf, epsfig}
\usepackage{color}
\usepackage{CJK} %chinese characters
\usepackage{amsmath}
\usepackage[normalem]{ulem}
\usepackage{bm}

\usepackage[colorlinks = true,
            linkcolor = blue,
            urlcolor  = blue,
            citecolor = blue,
            anchorcolor = blue]{hyperref}

\shorttitle{Dynamics of finite-size spheroids in turbulent flow}
\shortauthor{L. Jiang and other}

\title{Dynamics of finite-size spheroids in turbulent flow: the roles of flow structures and  particle boundary layers}
\author{Linfeng Jiang\aff{1}, Cheng Wang\aff{1}, Shuang Liu\aff{1},\\ 
	Chao Sun\aff{1,2}\corresp{\email{chaosun@tsinghua.edu.cn}},
	\and Enrico Calzavarini\aff{3}\corresp{\email{enrico.calzavarini@polytech-lille.fr}}}

\affiliation{\aff{1}Center for Combustion Energy, Key laboratory for Thermal Science and Power Engineering of Ministry of Education, Department of Energy and Power Engineering, Tsinghua University, Beijing 100084, China
	\aff{2}Department of Engineering Mechanics, School of Aerospace Engineering, Tsinghua University, Beijing 100084, China
	\aff{3}Univ.\ Lille, Unit\'e de M\'ecanique de Lille - J. Boussinesq - ULR 7512, F-59000 Lille, France}

\newcommand{\ec}[1]{\textcolor{black}{#1}}
\newcommand{\lj}[1]{\textcolor{black}{#1}}

\begin{document}
	
	\maketitle
	
	\begin{abstract}
		We study the translational and rotational dynamics of neutrally-buoyant finite-size spheroids in 
		hydrodynamic turbulence by means of fully resolved numerical simulations. We examine axisymmetric shapes, from oblate to prolate, and the particle volume dependences. 
		We show that the accelerations and rotations experienced by non-spherical inertial-scale particles result from volume filtered fluid forces and torques, similar to spherical particles.
		 However, the particle orientations carry signatures of preferential alignments with the surrounding flow structures, which is reflected in distinct axial and lateral fluctuations for accelerations and rotation rates. The randomization of orientations does not occur even for particles with volume equivalent diameter size in the inertial range, here \ec{up to} $60 \eta$ at $Re_{\lambda}=120$. Additionally, we demonstrate that the role of fluid boundary layers around the particles cannot be neglected to reach a quantitative understanding of particle statistical dynamics, as they affect the intensities of 
angular velocities, and the relative importance of tumbling with respect to spinning rotations. This study brings to the fore the importance of inertial-scale flow structures in homogeneous and isotropic turbulence and their impacts on the transport of neutrally-buoyant \lj{bodies with size in the inertial range}.	 
	\end{abstract}
	
	
	\section{Introduction}\label{sec:Intro}
	The motion of neutrally-buoyant finite-size material particles in a turbulent flow is typically regarded as the result of turbulent flow fluctuations occurring at the scale of the particle. In other words, the particles are seen as probes of a coarse-grained turbulent field. This implies that both the scaling trends in the fluctuations of  accelerations and rotational velocities as a function of the particle size are interpreted in terms of the known scaling of structure functions of turbulence in the framework of Kolmogorov 1941 (K41)  phenomenological theory \citep{2001NatureParticle,voth_laporta_crawford_alexander_bodenschatz_2002,qureshi2007turbulent,Brown2009prl,calzavarini2009acceleration,homann_bec_2010}. Such a picture has even been extended to the case of non-spherical axisymmetric particles with various aspect ratios to elucidate the observed different scaling trends in tumbling and spinning rotation rates \citep{ParsaPRL2014,bordoloi_variano_2017,BounouaPRL2018,Kuperman2019prf,pujara_voth_variano_2019,OehmkePRF2021}.  
	This interpretation assumes that neutrally-buoyant material particles are only weakly coupled to the flow, so that the feedback effects (also referred as two-way coupling)  stemming from the formation of boundary layers and wakes around the particle do not have a significant role, at least in statistical terms.\\
	This line of reasoning is markedly distinct from the one that is used to grasp the dynamics of small, i.e. sub- or Kolmogorov-scale ($\eta$) sized particles, where instead it is crucial to take into account  the local properties of the flow, notably the fluid acceleration and the velocity gradient and its topological properties along fluid Lagrangian trajectories \citep{Chong1990A,BenziPRE2009} . It is key here to recognize that \ec{distinctive} flow structures exist at small scales, such as the filament-like vortices where fluid tracers get trapped \citep{biferale2005trap,bentkamp2019persistent} or from which tiny inertial particles are ejected \citep{bec2007heavy} and to which spheroidal particles align \citep{PumirWilkinson2011, Gustavsson2014,ni_ouellette_voth_2014}.
	These two pictures, the dissipative and the inertial-scale ones, are somehow conceptually disconnected, with the only bridging idea being the call into play of so-called Fax\'en corrections which account for effects induced by the local curvature of the flow although at scales where the flow field is smooth, typically $\leq 10 \eta$, hence not yet in the inertial range \citep{calzavarini2009acceleration} \ec{(We note that while simple for spherical particles the Fax\'en corrections are more complex for other particle shapes, see \cite{dolata_zia_2021})}.\\ 
	In this paper, we aim at ameliorating the above described conceptual models by
	demonstrating that the dynamics of neutrally-buoyant inertial-sized particles: i) is affected by inertial-range coherent flow structures  with similar effects as the ones produced by dissipative-scale turbulent structures on small particles, and
	ii) is significantly 
	\ec{impacted}  by the particle feedback on the flow, especially if the considered observable is the angular rotation.
	This is here achieved by means of a novel series of 
	numerical experiments.
	First, we perform extensive numerical simulations of fully-resolved homogeneous isotropic turbulence (HIT) seeded with neutrally-buoyant spheroidal particles with sizes ranging from the upper bound of the dissipative range ($\sim 10\eta$) to the far inertial range ($60 \eta$). Second, we conduct a sequence of numerical experiments in the same fluid flow setting but with virtual particles, i.e.~particle toy models, where we vary the type or/and level of coupling between the flow and the particles.
	As we will show such virtual particles reveals to be a crucial tool for an original physical interpretation of the statistical properties of translation and rotation of the real material particles in interaction with the turbulent flow environment.\\
	The article is organized as follows. In the next section, we will describe in detail the dynamical  governing equations for the model system of particles in turbulence and its numerical implementation. We will dedicate a section to the description of the numerical experiments performed both with realistic and virtual particles. We later present the results on the particle acceleration and rotation rate statistics. In each of these two parts, we will guide the reader through a physical interpretation of the results that is mainly done by contrasting the simulations of real particles with virtual ones. Finally, the conclusions and perspectives opened-up by the study are given.
	
	\begin{table}
		\begin{center}
		\ec{
			\begin{tabular}{p{1.2cm}p{1.2cm}p{1.2cm}p{1.2cm}p{1.2cm}p{1.2cm}p{1.2cm}}
				$N^3$  & $\eta/\Delta x$ & $\tau_\eta/\Delta t$ & $L/\eta$ & $T_L/\tau_\eta$ & $\lambda/\eta$ & $Re_\lambda$\\
				$512^3$ &  $1.5$ & 230 & 175 & 47 & 22 & 120
			\end{tabular}
			\caption{Parameter of the NSE numerical simulations and relevant turbulence scales. $N^3$: number of spatial grid points, $\eta=(\nu^3/\epsilon)^{1/4}$: Kolmogorov dissipation length scale in grid space units $\Delta x$, $\tau_\eta$: Kolmogorov time scale in time-step units $\Delta t$, $L=u'^3/\epsilon$: integral scale, $T_L=L/u'$: large-eddy turnover time, $\lambda = (15 \nu u' / \epsilon )^{1/2}$: Taylor micro-scale, $Re_\lambda$: Taylor-Reynolds number.}
			}
			\label{table:NS}
		\end{center}
	\end{table}
	
	\begin{figure}
		\begin{center}
			\begin{minipage}[l]{0.6\columnwidth}
			\includegraphics[width=1\columnwidth]{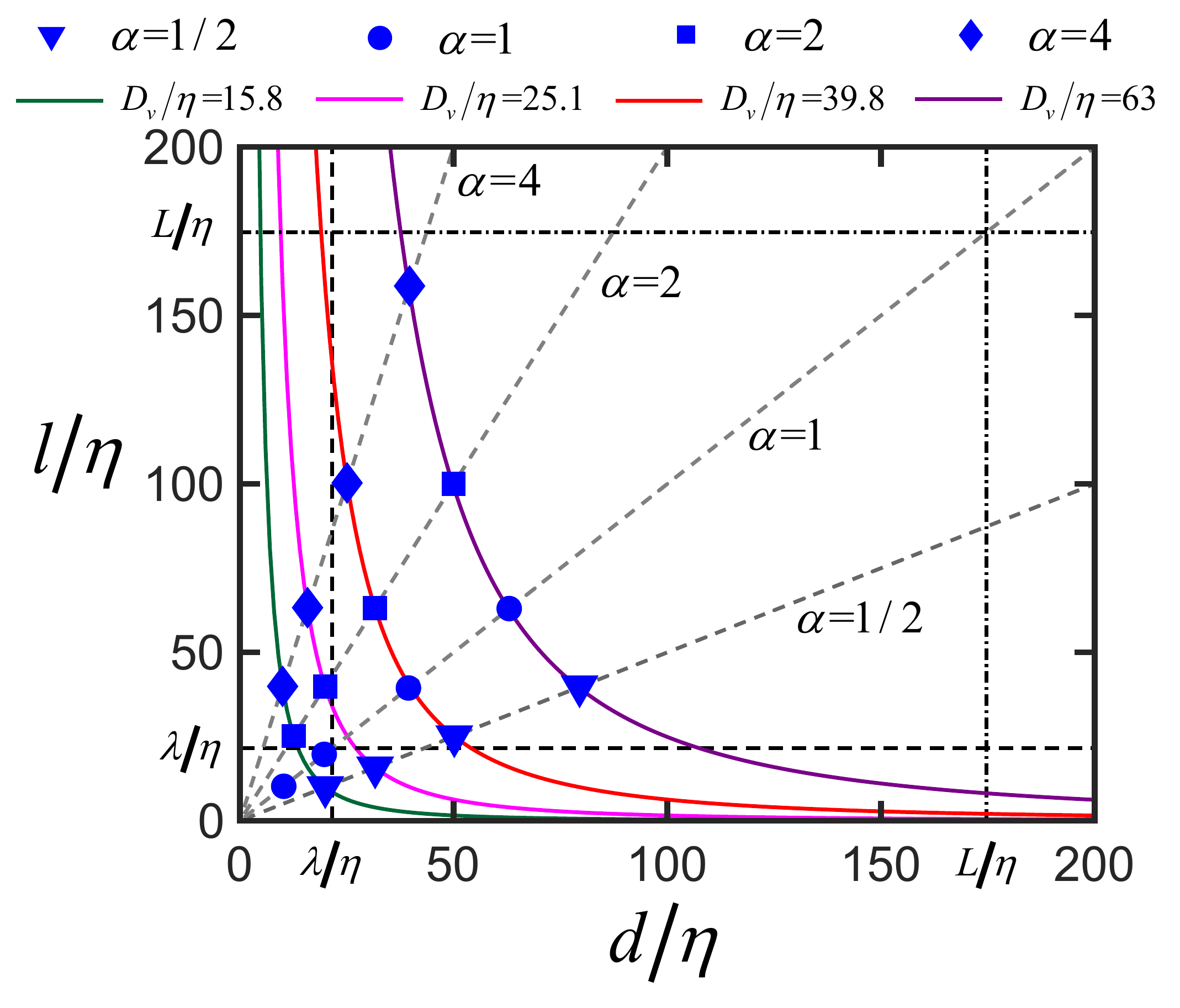}
			\end{minipage}
			\ec{
			$\quad$
			\begin{tabular}{|c|c|c|c|}
			\hline
			     $\alpha$ & $\frac{l}{\eta}$ & $\frac{d}{\eta}$ & $\frac{D_v}{\eta}$ \\
			     \hline
			     1/2 & 9.95 & 19.9 & 15.8\\
			     1/2 & 15.8 & 31.6 & 25.1\\
			     1/2 & 25 & 50.1 & 39.8\\
			     1/2 & 39.7 & 79.4 & 63\\
			     \hline
			     1 &10.2 &10.2 &  10.2 \\
			     1 & 19.7& 19.7 & 19.7\\
			     1 & 39.8 & 39.8 & 39.8\\
			     1 & 63 & 63 & 63\\
			     \hline
			     2 & 25.1 & 12.5 & 15.8\\
			     2 & 39.8 & 19.9 & 25.1\\
			     2 & 63.2  & 31.6 & 39.8\\
			     2 & 100 & 50 & 63\\
			     \hline
			     4 & 39.8 & 9.95 & 15.8\\
			     4 & 63.25 & 15.8 & 25.1\\
			     4 & 100.3 & 25.1 & 39.8\\
			     4 & 158.75 & 39.7 & 63\\
			     \hline
			\end{tabular}
			}
			\caption{Parameter space of particle shapes and sizes in turbulent scale units at $Re_\lambda=120$. The symbols indicate the numerically simulated particle types. All particles are axisymmetric of diameter $d$  and length $l$, aspect ratio $\alpha=l/d$, with $\alpha = 1/2,1,2,4$, the volume-equivalent-diameter is $D_v = (l d^2)^{1/3}$. \ec{The lengths are here all expressed in terms of the dissipative scale $\eta$. The Taylor micro-scale $\lambda$ (dashed lines) and  the integral scale $L$ (dashed-dotted lines) are also reported.}
			}
			\label{fig:fig_parameter}			
		\end{center}
	\end{figure}
	
	\section{Methods}
	
	We first describe the basic physical model system and the equation of motion ruling the dynamics of particles in a fluid flow. The numerical methods adopted and details on their implementations are very briefly outlined. 
	
	\subsection{The particles in turbulence model system and its numerical implementation}
	The fluid flow by which particles are transported is modeled by the incompressible Navier-Stokes equations (NSE) driven by an external random large-scale statistically homogeneous and isotropic force with constant in-time global energy input. This reads:
	\begin{eqnarray}
		\partial_t \textbf{u}  + \textbf{u}\cdot {\nabla}   \textbf{u} &=& -   \rho^{-1}{\nabla} p + \nu\ \nabla^2   \textbf{u} + \textbf{f}, \label{eq:N-S}\\
		{\nabla}  \cdot  \textbf{u} &=& 0, \label{eq:div1}
	\end{eqnarray}
	where $\textbf{u}(\textbf{x},t)$ denotes the fluid velocity vector field, $p(\textbf{x},t)$ is the hydrodynamic pressure, and the parameters are the kinematic viscosity $\nu$ and the reference liquid density $\rho$. \ec{The vector field $\textbf{f}$ refers  to the external force sustaining the HIT flow}.
	The particle-free turbulent flow intensity is here identified by a single dimensionless parameter, the Reynolds number based on the Taylor microscale, $Re_\lambda=\lambda u'/\nu$, where $u'=\sqrt{\langle u_iu_i\rangle^{v,t}/3}$ is the root-mean-square (r.m.s.) velocity of the turbulent flow ($\langle \ldots \rangle^{v,t}$ denotes here volume and time average), $\lambda=\sqrt{15\nu u'^2/\epsilon}$ is the Taylor length scale and $\epsilon = (\nu/2) \Sigma_{i,j}\langle(\nabla_i u_j + \nabla_j u_i)^2 \rangle^{v,t}$ is the mean global energy dissipation rate.
	
	The translation and rotation of rigid-body material particle is governed by the Newton-Euler equations (NEE):
	\begin{eqnarray}
		m_p \frac{\rm{d} \textbf{v}}{{\rm d}t} &=& \textbf{F}+\textbf{F}_c, \label{eq:Newton-Euler1}\\
		\frac{{\rm d}  \bm{\mathcal{I}} \bm{\Omega}}{{\rm d}t} &=& \textbf{T}+\textbf{T}_c, \label{eq:Newton-Euler2}
	\end{eqnarray}
	where $\textbf{v}(t)=d\textbf{r}/dt$ and $\bm{\Omega}(t)$ are the particle velocity and angular velocity vectors of a particle at position $\textbf{r}(t)$ with mass $m_p=\rho_p V_p$ ($\rho_p$ the particle density and $V_p$ the volume) and $\bm{\mathcal{I}}$ the moment of inertia tensor. 
Note that since we consider  neutrally buoyant and homogeneous particles, $\rho_p =\rho$, this allows neglecting the buoyancy force in the particle equation of motion as well any gravity-induced torque. Hence 
	$\textbf{F}$ and $\textbf{T}$ in (\ref{eq:Newton-Euler1})-(\ref{eq:Newton-Euler2}) denote here the hydrodynamic force and torque acting on the particle, 
	\ec{which are formally written as:
	\begin{eqnarray}
	    \textbf{F} &=& \oint_{S_p} \bm{\sigma} \cdot \textbf{n}\ dS \\
	    \textbf{T} &=& \oint_{S_p} (\textbf{x}-\textbf{r}) \times  (\bm{\sigma} \cdot \textbf{n})\ dS,
	\end{eqnarray}
	where $\bm{\sigma} = -p \bm{I}+\rho \nu (\nabla\textbf{u} +  \nabla\textbf{u}^T)$ is the fluid stress tensor, $\textbf{x}-\textbf{r}$ the position vector relative to the particle center and $\textbf{n}$ the outward-pointing normal to the particle surface $S_p$.}
	While $\textbf{F}_c$ and $\textbf{T}_c$ are the additional impulsive forces and torques related to the particle-particle collisions (the so-called four-way interaction).\ec{The two-way coupling, meaning the coupling between the fluid and the particles is provided by the requirement of no-slip boundary condition on the particle surfaces (see below for its numerical implementation).}
	
	For an axisymmetric particle with symmetry axis identified by a unit vector $\textbf{p}$, as the ones we consider in this study, the angular velocity $\bm \Omega$ can be decomposed into the tumbling rotation rate, ${\bm \dot{\textbf{p}}}={\bm\Omega}\times \textbf{p}$, the spinning rotation rate, $\bm\Omega^s=({\bm\Omega}\cdot \textbf{p})\textbf{p}$ \citep{VothARFM2017}. These will be key quantities in our analysis of particle angular dynamics. Similarly, it will reveal useful in our analysis to distinguish between axial and lateral accelerations, $\bm a = \rm {d}\textbf{v}/\rm{d}t$, denoted $\bm a^{\parallel}=(\bm a \cdot \textbf{p})\textbf{p}$ and $\bm a^{\perp} = \bm a\times\textbf{p}$ respectively.
	The shape of an axisymmetric spheroid is characterized by the aspect ratio, $\alpha=l/d$, where $l$ and $d$ are the sizes of the symmetry axis and of the one perpendicular to it. We will use the volume equivalent diameter $D_v=(d^2 l)^{1/3}=d\ \alpha^{1/3}$ as a parameter to compare the size of particles with different shapes.
	
	On the computational side: the NSE turbulent dynamics is here numerically simulated by means of a Lattice Boltzmann Method (LBM) code, the {\sc ch4-project}  \citep{Calzavarini_SI2019}
	which has been already extensively employed in studies of Lagrangian tracers and point-like particle dynamics in turbulence \citep{mathai2016microbubbles,calzavarini2020anisotropic,jiang2020rotation,jiang2021rotational}. The code uses a tri-linear scheme for the Lagrangian-Eulerian frame interpolations. The computational domain is cubic with equispaced grid sizes, $128^3$ and $512^3$, corresponding to $Re_\lambda=32,120$ \ec{(see Table \ref{table:NS} for details on numerical accuracy properties and physical characteristics of the simulated turbulent flow at with largest $Re_{\lambda}$)}.
	The NEE are numerically integrated with a second-order Adams-Bashforth time-stepping scheme. The particle-fluid two-way coupling is implemented by means of the immersed boundary method (IBM), which has been widely used in particle-laden flows \citep{peskin_2002,mittal2005immersed,uhlmann2005,luca2016IJMF,WangPRE2021} also in combination to LBM e.g. in \citep{SUZUKI2011173,Quang2014PRE}. \ec{The IBM enforce the no-penetration and no-slip boundary conditions at the fluid-particle interface by means of a localized feedback force, $\textbf{f}_p$, added to the NSE (\ref{eq:N-S}). Such $\textbf{f}_p$ term is also denoted as two-way coupling}. In order to ensure high-accuracy for the implementation of the no-slip fluid  boundary condition at the particle surface, we adopt the so-called IBM multi-forcing scheme with 5 step iterations, see  \citep{PRE_multiforce2007,Cheng2021JHD}. The non-spherical particle-particle interactions are implemented by means of soft-sphere collision forces \citep{PRE2015collision,luca2016IJMF} and lubrication corrections \citep{Brenner1961,cooley1969,PRE2015collision,luca2016IJMF} 
	In order to limit the effect of the pair interactions on the statistics of particle dynamics, the concentration of particles is kept very low (the maximal volume fraction is $1\%$) and only up to 4 particles are seeded in the simulation for the smallest particle size. All inter-particle collision events are removed from the data sets employed for the statistical analysis performed in the present study. \ec{To achieve this, the particle acceleration/rotation data around the collisions are removed by a time  window whose width is of the order of the correlation time of the particle acceleration (i.e. $ \gtrsim \tau_{\eta}$, see also Fig. \ref{fig:sm_validation}(b)).} Thus, the effect of pair interaction is considered negligible. 
	
	Prolate ($\alpha=2, 4$), spherical ($\alpha=1$), and oblate ($\alpha=0.5$) particles are investigated in the simulations, see Figure \ref{fig:fig_parameter} for a representation of the explored parameter space.  For all considered cases, the minimal linear size of the particle, either $d$ and $l$, is \ec{$> 15 \Delta x$} (grid units). \ec{Such choice guarantees a reasonable resolution of the particle boundary layers whose inner (or viscous) thickness is expected to be of order $\eta$ (although it might become thinner for particles with $D_v\gg \eta$ as pointed out in \citep{cisse2013slipping}).}
	The particle Stokes number, defined as the ratio of the particle viscous response time ($\tau_p = D_v^2 / (12 \nu)$)  with respect to the characteristic flow scale at the particle scale (here the turbulent eddy-turnover time $\tau_{\ell} = \ell^{2/3}/\epsilon^{1/3}$), can be expressed as $St_{\ell}=\frac{1}{12}(\frac{D_v}{\eta})^{4/3}$ \citep{xu2008motion,FiniteClusterPRE2012}. It varies in the range 1.8 to 20.9.
	Finally, for each studied particle type-case  the duration simulation is more than 100 large-eddy turnover times.
	
	We have validated the numerical results by comparing the translational dynamic of spheres ($\alpha=1$) at $Re_\lambda=32$  with a previous reference study \citep{homann_bec_2010}. This check validates our acceleration statistics (see Appendix \ref{sec:validation} for details). The rotational dynamics of the spheroids is instead checked by solving the rotation rate of a prolate spheroid in a pure shear flow, a condition for which the Jeffery solution \citep{Jeffery1922} is available, \ec{and by reproducing the numerical experiments in \citep{SUZUKI2011173}, which include settling of ellipsoidal particles in a quiescent flow at finite Reynolds numbers}.

	\begin{figure}
		\begin{center}
			\includegraphics[width=0.8\columnwidth]{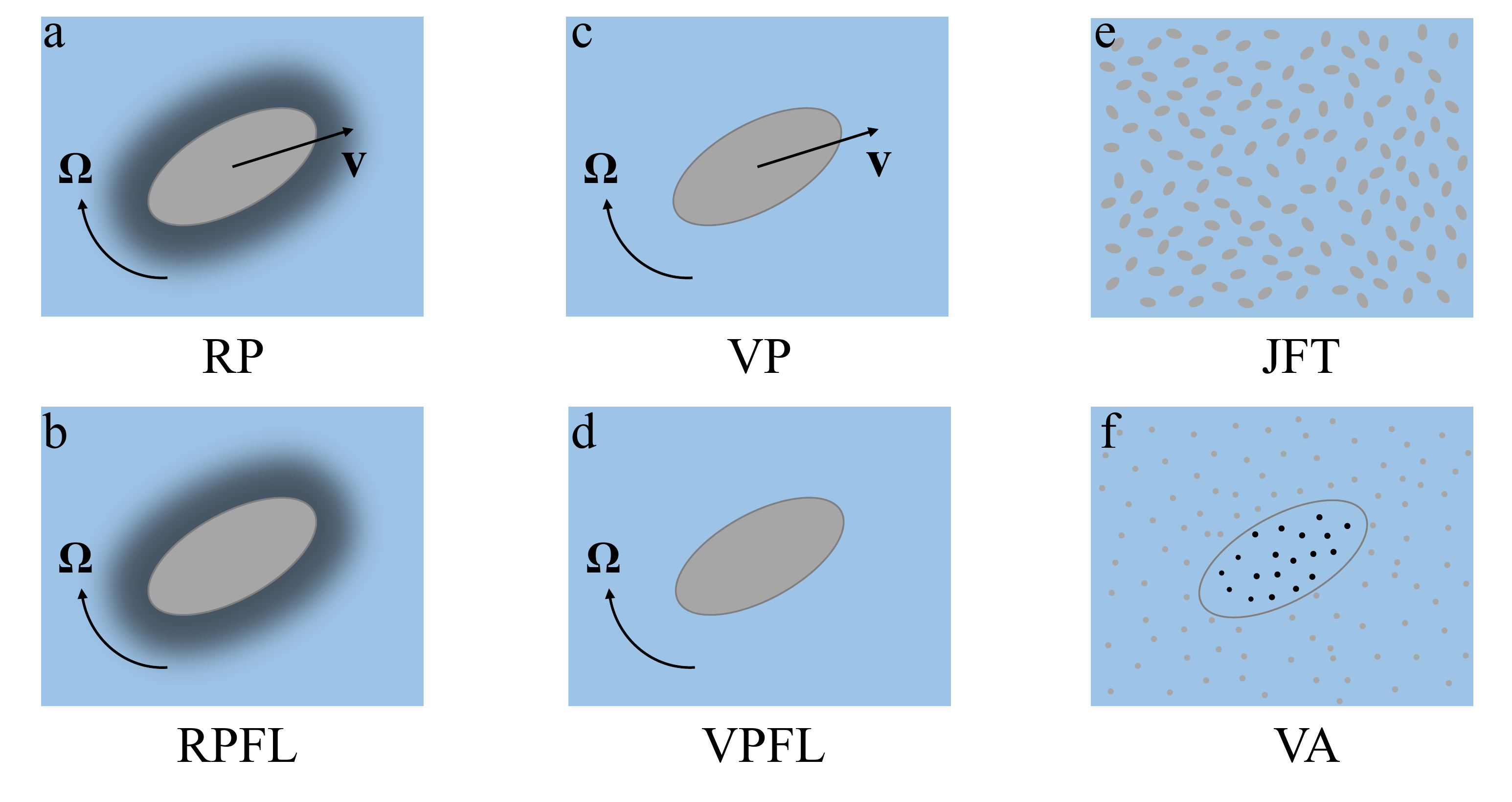}	
			\caption{Sketches of the six numerical experiments performed: (a) real particles RP, (b) real particles with fixed locations RPFL, (c) virtual particles VP, (d) virtual particles with fixed locations VPFL, (e) Jeffery fluid tracers JFT, and (f) volume average of Jeffery fluid tracers VA . The light blue region denotes the fluid. The black shaded areas in panels (a) and (b) represent the two-way coupling region. The points in panels (e) and (f) represent the Jeffery fluid tracers. The tracers marked in black in panel (f) are used to calculate the volume averages. 			}
			\label{fig:fig_numeric_exp}			
		\end{center}
	\end{figure}	
	
	%\textcolor{cyan}{
	\subsection{Numerical experiments with real and virtual particles}
	Taking advantage of the numerical approach, which allows to easily modify the equations of motions of the advected particles and so to create toy-models, we carry out several additional numerical experiments. This series of numerical experiments is described in the following.

	The first numerical experiment is the simulation of real particles (RP) which obeys the model system introduced in the previous section. 
	Their dynamics is governed by the NSE and NEE with fluid-solid coupling computed by the IBM algorithm, see Tab. \ref{table}, as also sketched in the cartoon of Figure~\ref{fig:fig_numeric_exp}a.
	
	The second numerical experiment consists of  spheroidal volume averages (VA) of fluid flow, as shown in Figure~\ref{fig:fig_numeric_exp}f. We average the fluid acceleration and vorticity over many ($N$) spheroids with the same diameter $D_v$ and aspect ratio $\alpha$ as the ones used for RP. 
This reads:	
%	\begin{eqnarray}
%		\langle a_i^2\rangle^{VA} &=& \frac{1}{N}\sum_{n=0}^{N}\left(\frac{1}{V^\alpha_{D_v}}\int_{V^\alpha_{D_v}} a_i d^3~\bm r\right)^2\\
%		\langle \Omega_i^2\rangle^{VA} &=& \frac{1}{N}\sum_{n=0}^{N}\left(\frac{1}{V^\alpha_{D_v}}\int_{V^\alpha_{D_v}} \omega_i d^3~\bm r\right)^2
%		\label{eq:virtualparticle}
%	\end{eqnarray}
	\begin{eqnarray}
		\langle a_i^2\rangle^{VA} &=& \frac{1}{N}\sum_{n=0}^{N}\left(\frac{1}{V_p}\int_{V_p} \frac{D u_i}{Dt}\ d^3 x\right)^2\\
		\langle \Omega_i^2\rangle^{VA} &=& \frac{1}{N}\sum_{n=0}^{N}\left(\frac{1}{V_p}\int_{V_p} \omega_i\ d^3 x\right)^2
		\label{eq:virtualparticle}
	\end{eqnarray}
	where $D u_i / Dt = \partial_t u_i + u_j \partial_j u_i$ is the i-th acceleration component the fluid inside a spheroid of volume  $V_p$.
	We note that the orientation and location of VA are fixed. However, in a statistical homogeneous and isotropic flow, as it is here, this is identical to performing the averages at locations random-uniformly distributed in space locations and over randomly oriented spheroids.

	In the third numerical experiment, we seed virtual particles (VP) in the HIT and evolve the trajectory and rotation by solving the NEE, as shown in Figure~\ref{fig:fig_numeric_exp}c. For these particles, the IBM algorithm in not activated and no feedback is implemented on the flow by the virtual particles ($\textbf{f}_p=0$). Thus, the translation and rotation of the particles are driven by the time derivative of the momentum and angular momentum of the fluid inside the virtual particles, which means the dynamics of VP is controlled by the fluid at the scale of VP. More explicitly the force\ec{, $\textbf{F}_{VP}$,} and torque\ec{, $\textbf{T}_{VP}$,} on the virtual particles are: 
	\ec{
	\begin{eqnarray}
		\textbf{F}_{VP} &=& \rho V_p \left< \frac{D \textbf{u}}{Dt} \right>^v \equiv \int_{V_p} \rho \frac{D \textbf{u}}{Dt}  ~ d^3  x\\
		\textbf{T}_{VP} &=& V_p \left< \frac{d\textbf{L}}{dt} \right>^v \equiv \int_{V_p}  (\bm x - \bm r)  \times \rho \frac{D \textbf{u}}{Dt}  ~ d^3 x	
	\end{eqnarray}
	}
	where $D \textbf{u}/Dt$ is the fluid acceleration, $\textbf{L}$ is the angular momentum, and $\langle\rangle^v$ denotes the average over the fluid volume of the virtual particle.
	
	In the fourth numerical experiment, we 
	adopt real particles with fixed locations (RPFL). We fix the spatial position of the real %\sout{particle} 
	particles and only evolve the rotation by solving the Eq.~(\ref{eq:Newton-Euler2}). In this case, the IBM is still implemented and the particles are two-way coupled with the flow, as shown in Figure~\ref{fig:fig_numeric_exp}b. 
	
	The fifth numerical experiment contains the virtual particles with fixed locations (VPFL), as shown in Figure~\ref{fig:fig_numeric_exp}d. We fixed the location of VP in the turbulence and only evolve the rotation by solving the Eq.~(\ref{eq:Newton-Euler2}). No feedback on the flow is imposed and the rotation of VPFL is driven by the time derivative of the  angular momentum of the fluid inside the virtual particles.
	
	Finally, to better contrast the dynamics of finite-size particles with respect to sub-Kolmogorov scale particles, we carry out a sixth numerical experiment with point-like spheroidal tracers, as shown in Figure~\ref{fig:fig_numeric_exp}e. The governing equations for the point-like spheroidal tracers with position, $\textbf{r}(t)$, and orientation, $\textbf{p}(t)$, are given by
	\begin{eqnarray}
		\dot{\textbf{r}} &=& \textbf{u}(\textbf{r}(t),t),\label{eq:part}\\
		\dot{\textbf{p}} &=& \bm \Omega \times \textbf{p}\label{eq:rot}\\
		{\bm \Omega} &=&\frac{1}{2}{\bm \omega}(\textbf{r}(t),t) + \Lambda\ \textbf{p}\times \mathcal{S}(\textbf{r}(t),t)\textbf{p}\label{eq:Jeffery3d}
	\end{eqnarray}
	Here, $\textbf{u}(\textbf{r}(t),t)$ is the fluid velocity at the particle position at time $t$. The vector ${\bm \omega}(\textbf{r}(t),t) = \nabla \times \textbf{u}$ is the fluid vorticity and $\mathcal{S} = ({\nabla}   \textbf{u}+ {\nabla}   \textbf{u}^T)/2$ is the strain-rate matrix of the fluid velocity gradient tensor, ${\nabla}   \textbf{u}$, at the particle position, and $\Lambda=\frac{\alpha^2-1}{\alpha^2+1}$ is the shape parameter of particles. The equations of rotation Eq.~(\ref{eq:rot})(\ref{eq:Jeffery3d}), called the Jeffery angular equations \citep{Jeffery1922,Byron2015}, have been extensively used for the study of the dynamics of tiny spheroidal tracers \citep{ParsaPRL2012,Gustavsson2014,calzavarini2020anisotropic,jiang2021rotational,jiang2020rotation}. The aspect ratios of the point-like spheroidal tracers (hereafter called Jeffery tracers) are chosen equal to the ones of RP. For each aspect ratio, we seed $10^5$ Jeffery tracers in the flow to have converged statistics. The numerical condition of the flow is identical to the simulation of RP. We use $\langle \rangle^{tracer}$ to denote the temporal and ensemble average of Jeffery tracers. Note that the translational dynamics of the particles is the one of perfect Lagrangian tracers, Eq. (\ref{eq:part}), independently of their geometric aspect ratio (i.e. the drag force is here neglected). For this reason, we denote the particles as Jeffery fluid tracers (JFT). For better clarity all the model systems described above are summarized in Table ~\ref{table}.
	\begin{table}
		\begin{center}
			\def~{\hphantom{0}}
			\begin{tabular}{cll}
				\textbf{n.} &
				\textbf{Numerical Experiments}  & \textbf{Model Equations} \\[3pt]
				1 & Real Particles (RP)                                                               & NSE + NEE + IBM \\ 
				2 & Volume Averages (VA)                                                              & NSE        \\ 
				3 & Virtual Particles (VP)                                                            & NSE + NEE       \\ 
				4 & Real Particles with Fixed Locations (RPFL)                                         & NSE + \ \ EE  + IBM \\ 
				5 & Virtual Particles with Fixed Locations (VPFL) 							          & NSE + \ \ EE   \\   
				6 & Jeffery Fluid Tracers (JFT)                                                       & NSE    +   tracer  eq.+   Jeffery eq.    \\ 
			\end{tabular}
			\caption{Summary of the six performed numerical experiments with the corresponding involved dynamical equations: Navier-Stokes equation (NSE) (\ref{eq:N-S})-(\ref{eq:div1}), Newton-Euler equation (NEE) (\ref{eq:Newton-Euler1})-(\ref{eq:Newton-Euler2}); Euler equation (EE) (\ref{eq:Newton-Euler2}) only rotation for particles with fixed location; No-slip boundary conditions (i.e. two-way coupling) at particle-fluid interface implemented via the immersed boundary method (IBM); tracer equation (\ref{eq:part}), Jeffery equation (\ref{eq:rot})-(\ref{eq:Jeffery3d}).}
			\label{table}
		\end{center}
	\end{table}

	\begin{figure}
		\begin{center}
			\includegraphics[width=1\columnwidth]{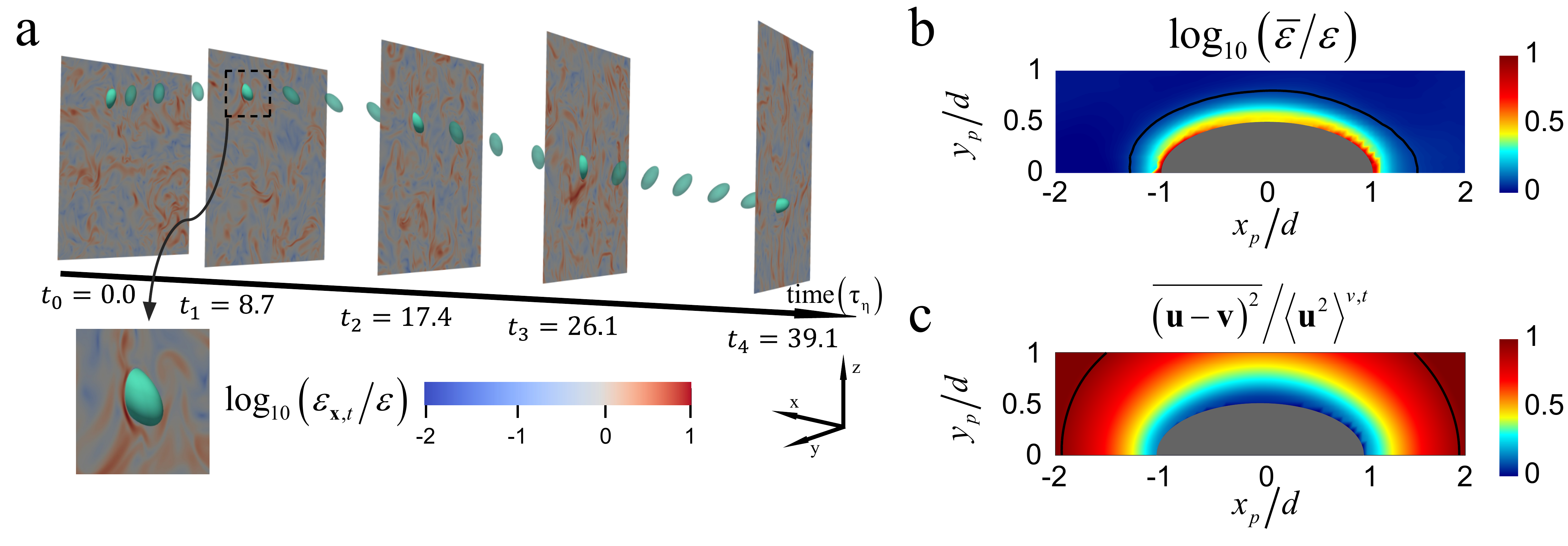}	
			\caption{
				(a) Example of trajectory for a spheroid with an equivalent diameter $D_v\approx 25~\eta$ and aspect ratio $\alpha=2$ at $Re_\lambda=120$. The slices show the local dissipation rate $\epsilon_{\textbf{x},t}$, normalized by $\epsilon$, at the planes through the spheroid center at time $t=0,8.7,17.4,26.1,39.1$ in $\tau_\eta$. 
				(b) Azimuthally and temporally averaged dissipation rate $\overline\epsilon$, normalized by $\epsilon$, around the spheroid with a diameter $D_v\approx 25~\eta$ and aspect ratio $\alpha=2$ at $Re_\lambda=120$. The black line denotes the contour line of $\rm log_{10}(\overline{\epsilon}/\epsilon)=0$. 
				(c) Azimuthally and temporally averaged square of  $\bm u-\bm v$, normalized by the mean square of $\bm u$, around the spheroid with diameter $D_v= 25~\eta$ and aspect ratio $\alpha=2$ at $Re_\lambda=120$. The black line denotes the contour line of $\overline{ (\bm u-\bm v)^2}/\langle \bm u^2\rangle^{v,t}=1$. $x_p$ and $y_p$ represent the coordinates in the particle frame. 
			}
			\label{fig:fig_visulization}			
		\end{center}
	\end{figure}
	\section{Results}
	\subsection{Particle trajectories and fields visualizations}
	In Figure~\ref{fig:fig_visulization}a a visualization example of the trajectory of a spheroid with $\alpha=2$ and an equivalent diameter $D_v=25 \eta$ over a relatively short time span, $\sim 40$ dissipative times ($\tau_{\eta}$), obtained from the simulation. The energy dissipation rate field in the fluid, $\epsilon$, is represented in colour over planes intersecting the spheroid. Such a quantity that depends on the square of velocity gradients allows to foreground the effect the fluid-particle coupling. We observe that just a thin layer of large dissipation appears around the particle (see the zoom-in region of the slice at $t_1=8.7\tau_\eta$), while no wakes are observed at larger distances. 
	This mild effect induced by the particle presence in the flow is better quantified by computing temporal fluid averages in the vicinity of the particle. 
	This is numerically performed by seeding the flow with point-like Lagrangian probes whose positions are fixed in the particle frame of reference. 
	%\sout{(see Appendix \ref{sec:outer_flow} for details)}. 
	Figure~\ref{fig:fig_visulization}b shows the temporally and azimuthally averaged local dissipation rate around the spheroid ($\alpha=2$ and $D_v/\eta=25$). It appears that the dissipation rate close to the particle is enhanced by one order of magnitude with respect to the far-field value, furthermore such an increase is uniformly distributed around the particle. Indeed, the contour line in the figure (black colour) marking where the particle influence reaches a saturation, indicates that the finite-size particle modifies the flow over the spheroidal region of about ~1.5 times of the linear size of the particle. Similar results for finite-size spheres have been observed in turbulence at $Re_\lambda=160$ \citep{cisse2013slipping}.
	The presence of the particle significantly modifies also the kinetic energy around the particle, the kinetic energy of the fluid is null at the particle interface and increases to the unperturbed value at about 
	twice the particle size $D_v$, see Figure~\ref{fig:fig_visulization}c. 
	\ec{The same figure suggests that the slip velocity of neutrally buoyant particles is small,  in particular $|\bm{u}-\bm{v}| \sim \bm{u}_{rms}$.
    This implies that the particle Reynolds number $Re_p = \langle |\bm{u}-\bm{v}| \rangle D_v / \nu \sim (D_v/\lambda) Re_{\lambda} \sim O(10^2)$.
    If the flow surrounding the particle were steady and laminar this would lead to a steady or periodically oscillating wake. In a turbulent flow, where the direction of motion changes frequently, such an effect is weakened. For the cases of spherical particles, this agrees with detailed study by \cite{cisse2013slipping}.}
	Overall the presence of a neutrally-buoyant finite-sized non-spherical particle has weak feedback on the turbulent flow field. This is certainly very specific to the case of inclusions with the same density, the situation is very different in the case of heavy particles \citep{BRANDLEDEMOTTA2016369} or large bubbles \citep{mathai2018dispersion}. This suggests that the picture of neutral inertial particles as essentially passive objects might be sound. We will now address this in more detail by looking at two quantities that are most sensitive to the fluid-particle one-way coupling, i.e., the fluid acceleration and later the angular velocity.  
	
	\subsection{\ec{Particle Acceleration statistics:\\ the VP model explains the size and shape dependencies in RP}}
\begin{figure*}
		\begin{center}
			\includegraphics[width=1.0\columnwidth]{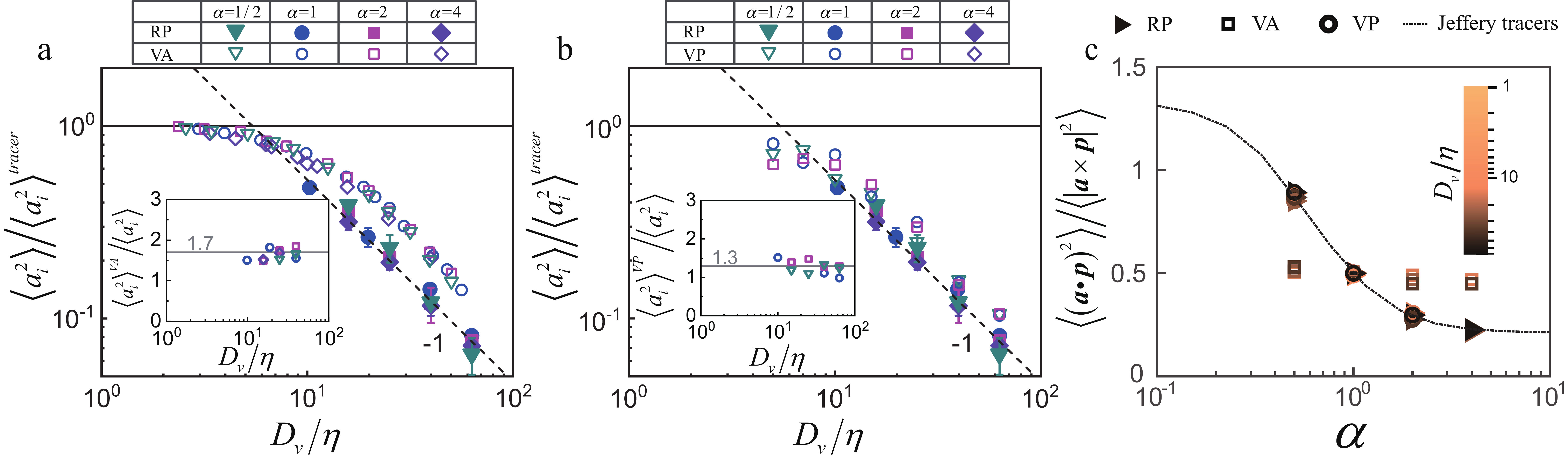}
			\caption{(a) Normalized single component acceleration variance as a function of the normalized particle size $D_v/\eta$ at $Re_\lambda=120$ for real particles (RP: solid symbols) and volume averages (VA: open symbols). The dashed lines are the fitting lines for spheres of RP (exponent $-1.0 \pm 0.1$). Inset in (a) shows the ratio of acceleration variance of VA to acceleration variance of RP, $\langle a_i^2 \rangle^{VA}/\langle a_i^2 \rangle$, as a function of $D_v/\eta$. (b) Normalized componentwise acceleration variance as a function of the normalized particle size $D_v/\eta$ at $Re_\lambda=120$ for real particles (RP: solid symbols) and virtual  particles (VP: open symbols). The dashed lines are the fitting lines for spheres of RP. Inset in (b) shows the ratio of acceleration variance of VP to acceleration variance of RP, $\langle a_i^2 \rangle^{VP}/\langle a_i^2 \rangle$, as a function of $D_v/\eta$. (c) The ratio of axial variance $\langle(\bm a\cdot\bm p)^2\rangle$ to lateral variance $\langle|\bm a\times\bm p|^2\rangle$ of particle acceleration as a function of aspect ratio at $Re_\lambda=120$ for RP (triangles), VA (square), and VP (circles), respectively. The black dash-dot line denotes the results of Jeffery tracers. The color represents the normalized particle size $D_v/\eta$ in the logarithmic scale.
			}
			\label{fig:VA_a_p}			
		\end{center}
	\end{figure*}

	%\subsection{Particle acceleration statistics}
	We begin discussing the dependency of the particle  acceleration intensity of large spheroids as a function of their shape and size and the physical mechanisms responsible for the observed dependencies.  
	Figure~\ref{fig:VA_a_p}a reports the temporal and ensemble average (denoted in the following $\langle \ldots \rangle$) of the single-cartesian-component acceleration variance $\langle a_i^2\rangle$ of real particles (RP), normalized by the acceleration variance of fluid tracers $\langle a_i^2\rangle^{tracer}$, as a function of the equivalent diameter in turbulent dissipative scale units $D_v/\eta$. % at  $Re_\lambda=120$.
	Two observations are here in order. First, the acceleration variance decreases as the size of the spheroids increases. In the range $D_v/\eta\in[10,63]$ we measure power-law \ec{$\sim(D_v/\eta)^{-1.0 \pm 0.1}$} (at  $Re_\lambda=120$). 
	Second, we remark a good collapse 
	onto a single curve of all the data points corresponding to different aspect ratios (with $\alpha= [1/2,1,2,4]$). 
	The observed scaling behaviour is consistent with the one observed for spherical particles at similar $Re_\lambda$ values \citep{volk2011dynamics}. 
	Such a scaling deviates from the $(D_v/\eta)^{-2/3}$ that can be guessed on the basis of dimensional reasoning relying on the Kolmogorov (K41) theory of turbulence  \citep{ voth_laporta_crawford_alexander_bodenschatz_2002, qureshi2007turbulent}.  
	This reasoning assumes that the particles are only one-way coupled to the fluid, i.e., they do not affect the fluid flow significantly through particle-fluid interactions. 
	However, the existence of such a deviation  has been already reported in studies on spherical particles. In particular, it is known that the scaling exponent has a weak Reynolds number dependence \citep{voth_laporta_crawford_alexander_bodenschatz_2002,qureshi2007turbulent}.  At lower $Re_\lambda$ than the one used in this study the observed exponent is closer to the value 
	$-4/3 $ as observed in Ref.~\citep{bec2007heavy}. The role of the  pressure scaling dependence on  $Re_\lambda$ has been conjectured to be responsible for such a behaviour \citep{bec2007heavy}. Our results at low Reynolds number, $Re_\lambda=32$ (see Figure~\ref{fig:sm_validation} in Appendix \ref{sec:validation}) confirms such a dependence. 
	On the other hand,  the collapse of acceleration variance for different particle shapes (Figure~\ref{fig:VA_a_p}a) is a novel  feature. It is remarkable because it implies that the hydrodynamical forces involved in the translational dynamics of spheroids are overall determined by the particle volume.

In order to further substantiate these findings, we show the results of the numerical experiment with fluid volume averages (VA) and the so-called virtual particles (VP) in Figure \ref{fig:VA_a_p}(a and b) (empty symbols). Similar scaling relations of acceleration variances are observed both for VA and VP in the inertial regime, whereas the magnitudes of the acceleration variances of VP and VA are not equal. In particular, the amplitude of accelerations fluctuations of VP particles is closer to the ones of RP particles (see insets in Figure~\ref{fig:VA_a_p}(a and b) where their ratio is plotted).
Note that the acceleration variances of VA and VP represent respectively the volume averaging of the fluid flow acceleration in the fixed frame (Eulerian perspective) and in an advected frame (Lagrangian perspective). Therefore, these results stress that the scaling behaviour of the translational dynamics is dominated by the volume averaging instead of other surface forces such as e.g. the drag. 
However, an interesting question is: Why VP are capable to better approximate the statistics of real particles than VA? This might be due to preferential sampling of fluid structures along their Lagrangian trajectories, a fact that we tend to exclude because we know that these particles do not form clusters, or this might come from a preferential orientation in space (that is dynamical for the RP and VP cases while absent for VA). This is at odds with the expectations that particles of sizes well into the inertial range, shall become randomly oriented due to the averaging of turbulent fluid fluctuations that results from the spatial volume filtering.

%\subsubsection{\ec{The flow structures and the acceleration preferential alignment}}
In order to understand if any correlation exists between the particle acceleration and its orientation, we look at 
 the ratio of the particle axial acceleration variance $\langle(\bm a\cdot\bm p)^2\rangle$ to the lateral acceleration variance $\langle|\bm a\times\bm p|^2\rangle$ as a function of their aspect ratio, Figure~\ref{fig:VA_a_p}c. For spherical particles ($\alpha=1$), no preferential alignment is expected  between $\textbf{a}$ and $\textbf{p}$, even for various sizes, due to the full rotational symmetry of the body shape. For the non-spherical case, a value of the ratio different from 0.5 marks the existence of preferential orientation.

	\begin{figure}
		\begin{center}
			\includegraphics[width=0.7\columnwidth]{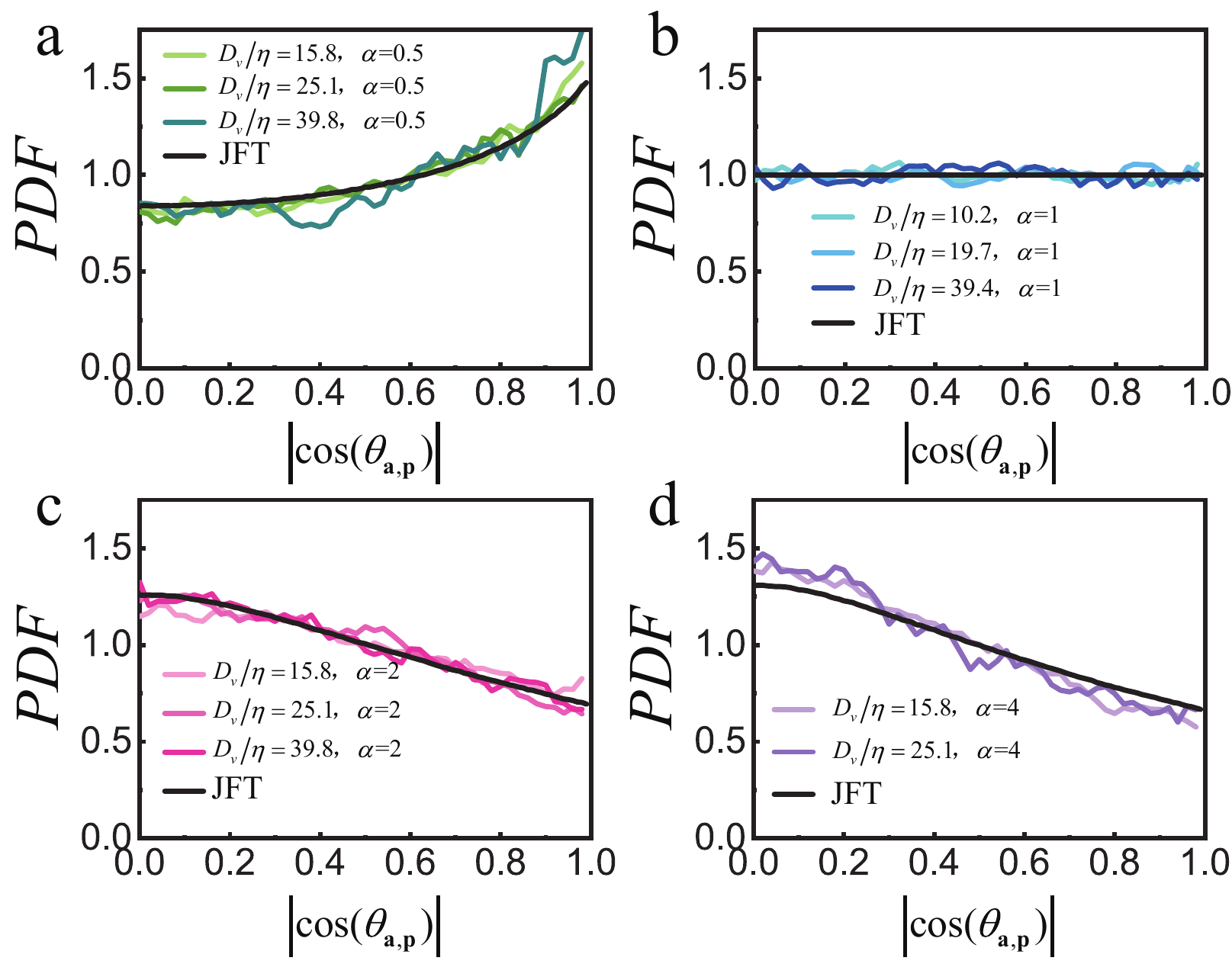}
			\caption{PDFs of $|\rm cos(\theta_{\textbf{a},\textbf{p}})|$ for RP spheroids with different aspect ratios: (a) $\alpha=0.5$, (b) $\alpha=1$, (c) $\alpha=2$, and (d) $\alpha=4$ and for Jeffery fluid tracers. 
			}
			\label{fig:SM_alignment}			
		\end{center}
	\end{figure}
	
	We find that the oblate particles preferentially align with their acceleration ($ \langle(\bm a\cdot\bm p)^2\rangle/\langle|\bm a\times\bm p|^2\rangle>0.5$) and the prolate particles are predominantly perpendicular to the acceleration ($ \langle(\bm a\cdot\bm p)^2\rangle/\langle|\bm a\times\bm p|^2\rangle<0.5$). It also turns out that the alignment of large particles is identical to that of Jeffery fluid tracers (JFT), which indicates that the alignment is insensitive to the particle's size. This is further confirmed by the measurements of the PDFs of $|cos(\theta_{\bm a, \bm p})|$ for various sizes and shapes (see Figure~\ref{fig:SM_alignment}), where $\theta_{ap}$ defines the angle between the acceleration vector $\bm a$ and orientation vector $\bm p$. The particular trend of $ \langle(\bm a\cdot\bm p)^2\rangle/\langle|\bm a\times\bm p|^2\rangle$
	as a function of $\alpha$ can then be understood in terms of the knowledge of Jeffery fluid tracers.
	It has been shown experimentally that prolate fluid tracers have accelerations preferentially perpendicular to the local vorticity \citep{Liberzon_acceleration2012}. It has also been theoretically demonstrated that prolate particles preferentially align with vorticity while oblate particles are preferentially perpendicular to it \citep{PumirWilkinson2011}.
	The combination of these two pieces of evidence explain qualitatively the observed alignment, depending on the aspect ratio, between the particle orientation and the acceleration.
	The existence of these alignments indicates that the translation of finite-size neutrally-buoyant particles in a flow is, in this respect, surprisingly similar to that of point-like inertialess axisymmetric fluid tracers, or conversely that geometrical structure of flow at the inertial scale shows a similar structure with the one of the velocity gradient, which occurs at the dissipation scale. 
	
	We now turn the attention to the results for VA and VP. For VA, the values of the ratio $\langle(\bm a\cdot\bm p)^2\rangle/\langle|\bm a\times\bm p|^2\rangle$ are close to 0.5 for all considered sizes and aspect ratios, which is expected because VA is not aligned with the corresponding acceleration vector. Remarkably, it is found that the ratio $\langle(\bm a\cdot\bm p)^2\rangle/\langle|\bm a\times\bm p|^2\rangle$ of VP shows excellent agreement with that of RP (Figure~\ref{fig:VA_a_p}c). This suggests that the effect induced by the two-way coupling 
	on the particle preferential alignment, if present, is weak. 
	The similarity between RP and VP implies that the neutrally-buoyant finite-size particle behaves roughly like a mass of fluid with the same particle size and shape. Furthermore, the size independence of the alignment indicate that the Lagrangian behavior of a fluid mass at the inertial scale is similar to the flow at the dissipative scale. The neutrally-buoyant finite-size particles can then be regarded as probes of the Lagrangian coarse-grained field of turbulence.
	
	In the next section, we will check if the similarity between real and virtual particles still holds when the rotational properties of particles are considered.\\
	\\
	\\
	\\
	\begin{figure}
		\begin{center}
			\includegraphics[width=0.9\columnwidth]{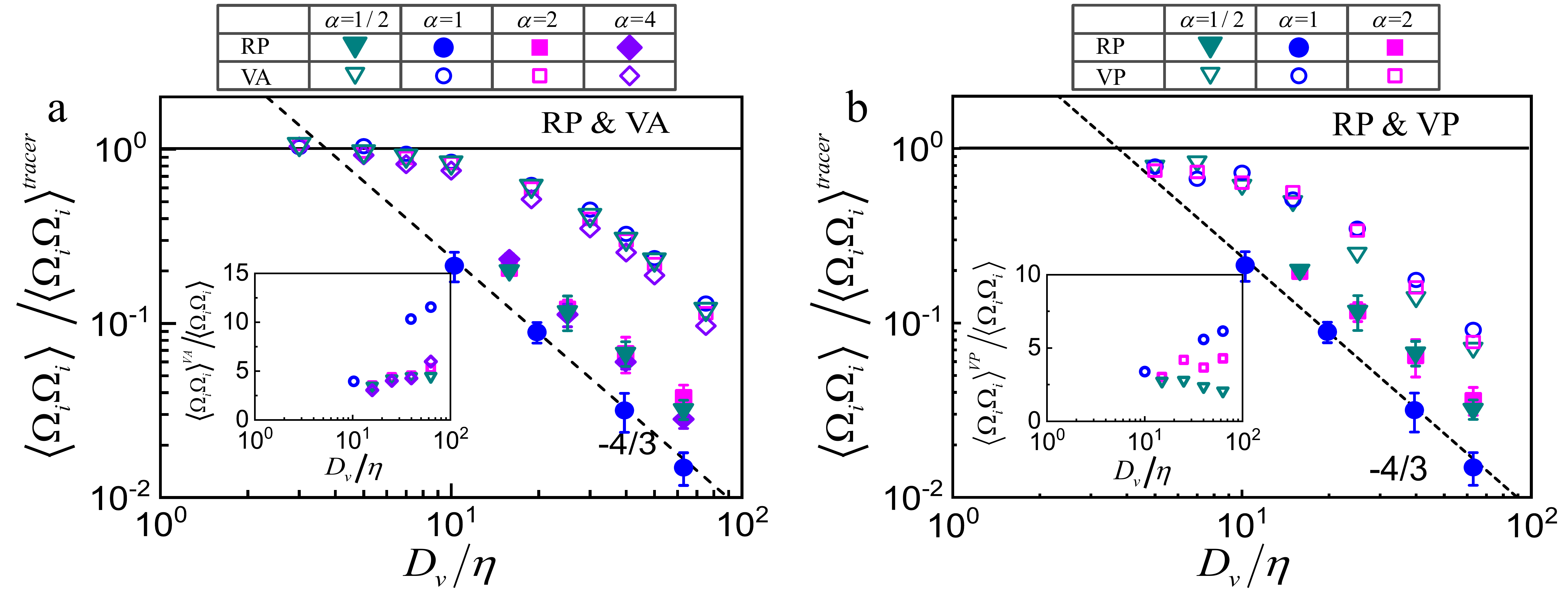}
			\caption{(a) Mean squared angular velocity, normalized by the results of Jeffery tracers with corresponding aspect ratio, as a function of $D_v/\eta$ at $Re_\lambda=120$. The open symbols denote the results of VA for spheres. The dashed line denotes the fitting line for spheres with scaling exponent -4/3. Inset shows the ratio of angular velocity variance of VA to that of RP, $\langle \Omega_i\Omega_i \rangle^{VA}/\langle \Omega_i\Omega_i \rangle$, as a function of $D_v/\eta$ for spheres. (b) Mean squared angular velocity, normalized by the results of Jeffery tracers with corresponding aspect ratio, as a function of $D_v/\eta$ at $Re_\lambda=120$. The open symbols denote the results of VP. The dashed line denotes the fitting line for spheres with scaling exponent -4/3. Inset shows the ratio of angular velocity variance of VP to that of RP, $\langle \Omega_i\Omega_i \rangle^{VP}/\langle \Omega_i\Omega_i \rangle$, as a function of $D_v/\eta$.
			}
			\label{fig:VA_rotation}			
		\end{center}
	\end{figure}	
	
	%\subsection{Particle rotation statistics}
	\subsection{\ec{Particle rotation statistics:\\ the VP model fails to describe the shape dependence in RP}}
	
	As already mentioned in Sec.~\ref{sec:Intro}, the rotations of tiny, $D_v \lesssim \eta$, neutrally-buoyant particles in turbulent flows are determined by the evolution of the fluid velocity gradient along the Lagrangian particle trajectories. The Jeffery equation (\ref{eq:rot})- (\ref{eq:Jeffery3d}) well captures the angular dynamics of such particles.  %\citep{Jeffery1922}.
	This has been recently proven in a series of combined experimental and numerical studies \citep{ParsaPRL2012,Byron2015,calzavarini2020anisotropic,jiang2021rotational}.
	When it comes to larger particles, in particular to inertial scale particles, the physics is more complicated due to: i) the non-smooth character of the flow field at the particle scale, ii) the spatio-temporal filtering of flow fluctuations associated with the particle size and inertia and iii) the two-way coupling with the fluid flow. This means that not only the shape (parametrized by $\alpha$) but also the particle sizes become relevant parameters of the problem.
	This is easily grasped in Figure~\ref{fig:VA_rotation}, where we show the mean square angular velocity, normalized by the value of the mean square angular velocity of point-like Jeffery fluid tracers (JFT) with corresponding aspect ratio, as a function of the normalized diameter $D_v/\eta$  for RP (solid symbols). As the size increases, we observe that the mean square angular velocity decreases and scales approximately as $\langle {\Omega}_i{\Omega}_i\rangle\sim (D_v/\eta)^{-4/3}$ for $D_v \geq 10 \eta$.
	Since the angular velocity of JFT is nearly constant at changing the aspect ratio \citep{Byron2015}, the decrease of angular velocity is exclusively an inertial term.
	Even if an offset between spherical and non-spherical particles is observed, the good collapse for prolate and oblate spheroids with different shapes substantiates the equivalent diameter $D_v$ as the relevant characteristic scale also for the rotational dynamics. We recall that the scaling exponent -4/3 can be dimensionally deduced using the relations of the K41 turbulence theory. Specifically by assuming that the rotation rate is related to the eddy turnover time at the particle scale, as confirmed in experiments with long slender fibers \citep{ParsaPRL2014,BounouaPRL2018} and large cylinders \citep{bordoloi_variano_2017,Pujara2018prf}.

	The VA and VP numerical experiments exhibit also a similar behaviour as RP, i.e. 
	a decreasing trend with $D_v$ consistent with the $-4/3$ power-law. However, the magnitude of the mean squared angular velocity is larger for VA and VP than that of RP (see again Figure~\ref{fig:VA_rotation}(a,b)). This suggests that rotational dynamics of finite-size neutrally-buoyant spheroids (RP) is also dominated by the volume filtering effect, as proven by its scaling-behavior with respect to $D_v$, while its amplitude must be affected by other physical effects (to be discussed later). 
	From the insets of Figure~\ref{fig:VA_rotation}, we note that the angular velocity variance discrepancy between VA and RP and VP and RP is larger than the discrepancy of acceleration variances.
In conclusion, as far as angular velocity is concerned we can not say that VPs are a good approximation of RP particles.  This discrepancy is even more important when one looks at the ratio between the intensity of axial as compared to lateral rotational components, that is to say, tumbling and spinning, as we will discuss in the following. 

	\begin{figure}
		\begin{center}
			\includegraphics[width=0.9\columnwidth]{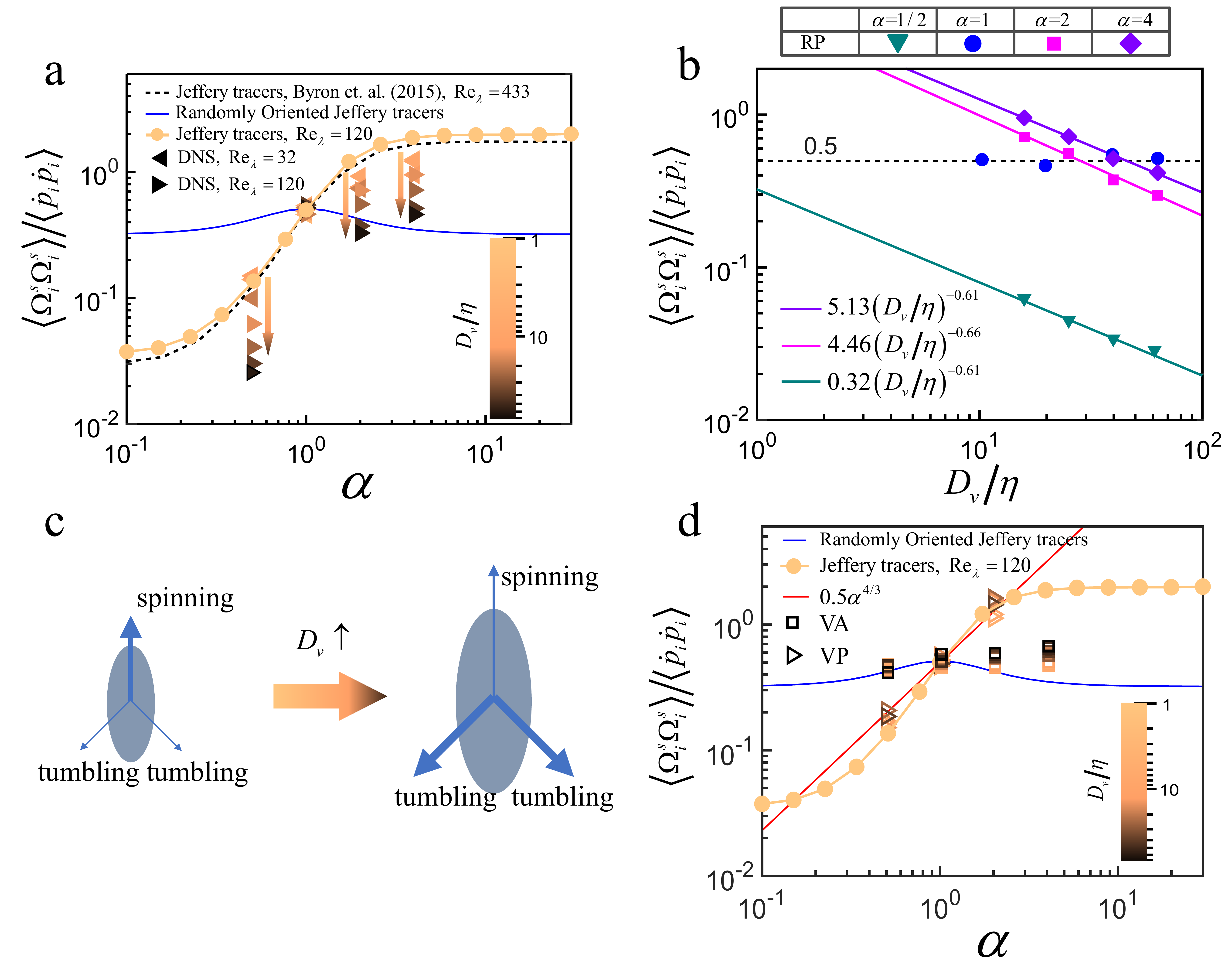}
			\caption{ 
			(a) The ratio of spinning rate squared to tumbling rate squared, $\langle \Omega_i^s\Omega_i^s\rangle /\langle \dot{p}_i\dot{p}_i\rangle$, as a function of aspect ratio $\alpha$ for RP. The color bar represents the normalized particle size $D_v/\eta$ in the logarithmic scale. Spheres denote the results for Jeffery tracers. Leftward triangles denote the results at $Re_\lambda=32$ and rightward triangles denote the results at $Re_\lambda=120$. The arrows denote the direction that the size of particles increases. The dashed line shows the results of tracers at $Re_\lambda=433$ in Ref.~\citep{Byron2015}. (b) The ratio of mean spinning rate squared to mean tumbling rate squared, $\langle {\Omega}_i^s{\Omega}_i^s\rangle/\langle\dot{p}_i\dot{p}_i\rangle$, as a function of size $D_v/\eta$.(c) The cartoon illustrates that the weight of the tumbling rate in the angular velocity of prolate spheroids becomes larger as the particle size increases.
			(d) The ratio of spinning rate squared to tumbling rate squared, $\langle \Omega_i^s\Omega_i^s\rangle /\langle \dot{p}_i\dot{p}_i\rangle$, as a function of aspect ratio $\alpha$ for VA and VP. The color bar represents the normalized particle size $D_v/\eta$ in the logarithmic scale.
			}
			\label{fig:VA_rotation_ratio}			
		\end{center}
	\end{figure}
	
In Figure~\ref{fig:VA_rotation_ratio}a, we show the ratio $\langle {\Omega}_i^s{\Omega}_i^s\rangle/\langle\dot{p}_i\dot{p}_i\rangle$ as a function of aspect ratio $\alpha$ for particles of different sizes. 
We first discuss the behaviour of sub-Kolomogorov scale particles, whose dynamics is well described by the Jeffery equation. In this case, if the particle orientation vector were uncorrelated with the fluid velocity gradients, and $\textbf{p}$ is an isotropic vector, upon time and ensemble averaging Eq. (\ref{eq:Jeffery3d}), one obtains:
	\begin{equation}
		\frac{\langle {\Omega}_i^s{\Omega}_i^s\rangle}{\langle\dot{p}_i\dot{p}_i\rangle}= \frac{ \frac{1}{12}\langle \omega^2 \rangle}{\frac{1}{6} \langle \omega^2 \rangle +   \frac{1}{5} \Lambda^2 \langle \mathcal{S}:\mathcal{S} \rangle} = \frac{5}{10 + 6 \Lambda^2}.
		\label{eq:rand_ratio}
	\end{equation}
	Note that the last equality follows from the statistical properties of the HIT flow, where $\frac{1}{2} \langle \omega^2 \rangle  = 2 \langle \mathcal{S}:\mathcal{S} \rangle = \epsilon/\nu$ \citep{ParsaPRL2012,Byron2015}.
	This would produce only a mild variation of  the spinning to tumbling ratio across the whole range of aspect ratios, as it is shown in  Figure~\ref{fig:VA_rotation_ratio}a (blue solid curve).
	However, it is known that small particles develop strong correlations with velocity gradients that are responsible for a sharp deviation from this prediction.
	The present Jeffery' models simulations 
	clearly show this behaviour in Figure~\ref{fig:VA_rotation_ratio}a (spheres symbols).
	Incidentally, we note that our results at $Re_\lambda=120$ agree with previous numerical study (dashed line) at $Re_\lambda=433$ \citep{Byron2015}, underlying the weak $Re_\lambda$ dependence of this phenomenon. For spheres, the angular velocity is isotropic and the ratio $\langle {\Omega}_i^s{\Omega}_i^s\rangle/\langle\dot{p}_i\dot{p}_i\rangle$ stays around 0.5 for all cases. A marked asymmetry exists between oblate and prolate particles, while the first tumble intensely the seconds are spin dominated, as they tend to align with the vorticity vector \citep{VothARFM2017,ParsaPRL2012}. Futhermore, a saturation of these behaviours is observed in both cases for extreme aspect ratios, i.e. $\alpha \lesssim 1/4$ and $\alpha \gtrsim 1/4$. 
	
	What happens when particles have a finite size? 
	A naive K41-based prediction would suggest that the tumbling scales with the length  $\langle\dot{p}_i\dot{p}_i\rangle \sim l^{-4/3}$ and that the spinning scale with the $\langle {\Omega}_i^s{\Omega}_i^s \rangle \sim 0.5 d^{-4/3}$ (where the 0.5 factor comes for the matching with the spherical case) so that
	\begin{equation}
	    \frac{\langle {\Omega}_i^s{\Omega}_i^s\rangle}{\langle\dot{p}_i\dot{p}_i\rangle} \sim \frac{\alpha^{4/3}}{2} .
	    \label{eq:k41_ratio}
	\end{equation}
	However, this prediction falls largely off the  result of the simulations, where we see a marked decreasing $D_v$ dependence of the ratio of the mean square spinning to tumbling rate (see Figure~\ref{fig:VA_rotation_ratio}a and b). Larger particles in proportion spin less and tumble more.  This is indeed true for all aspect ratios except for spheres (see Figure~\ref{fig:VA_rotation_ratio} d).
	It has been pointed out by \citep{OehmkePRF2021}, that these types of dimensional reasoning ignore the decorrelation of the velocity field along the particle surface. The rotation rates stem from the integral of the velocity gradients along with the particles and as such e.g. the spinning rate can be much attenuated in long fibers ($\alpha \gg 1$).
	This integral effect is possibly more important when both the diameter and the length are much larger than the dissipation scale.
	As a consequence, the scaling behaviors of the spinning and tumbling rate can deviate from K41-based scalings as another length scale is involved, and the ratio $\langle {\Omega}_i^s{\Omega}_i^s\rangle/\langle\dot{p}_i\dot{p}_i\rangle$ becomes to be a function of the particle size.
	
	The experiments by \cite{OehmkePRF2021}, the only to date to have measured both the spinning and tumbling of finite-size particles, find that for prolate fibers the variance of the spinning rate is always larger than the variance of the tumbling rate. These authors comment that the observed behaviour   ``surprisingly resembles that observed previously for sub-Kolmogorov fibers''. 
	These observations are fully confirmed by our simulations \ec{(see Appendix \ref{sec:comparison} for a more detailed comparison)}. However, we additionally observe a new feature:
	
	The ratio $\langle {\Omega}_i^s{\Omega}_i^s\rangle/\langle\dot{p}_i\dot{p}_i\rangle$ for oblate and prolate spheroids shows systematic decreasing trends as the particle equivalent size increases, and this  for both $Re_\lambda=32$ and $Re_\lambda=120$.

	For oblate spheroids ($\alpha=0.5$), the ratio becomes vanishingly small at increasing their size, which indicates that the oblate spheroids hardly spin when their size is large compared to the dissipation scale. Similarly, prolate spheroids of large size reduce their spinning to the point that it becomes smaller than tumbling ( Figure~\ref{fig:VA_rotation_ratio}a).
	
	Figure~\ref{fig:VA_rotation_ratio}b shows the ratio $\langle {\Omega}_i^s{\Omega}_i^s\rangle/\langle\dot{p}_i\dot{p}_i\rangle$ as a function of the equivalent particle size $D_v/\eta$. It is found that the ratio $\langle {\Omega}_i^s{\Omega}_i^s\rangle/\langle\dot{p}_i\dot{p}_i\rangle \sim(D_v/\eta)^{-0.6}$ for all considered shapes.  We remark that the existence of the observed scaling behavior of the ratio $\langle {\Omega}_i^s{\Omega}_i^s\rangle/\langle\dot{p}_i\dot{p}_i\rangle$ is not certain. However, this emphasizes that the scaling exponents of the mean square tumbling rate and spinning rate can not be both equal to -4/3. To our knowledge, no presently available model accounts for the observed behaviour.
	
	In summary, the main feature connected to the finite size is that larger particles tend to tumble much more than spinning (see cartoon Figure~\ref{fig:VA_rotation_ratio}c).
	Further, we observe that the VA and VP measurements do not have this property. While VA nearly keeps the value of the Jeffery randomly oriented prediction, VPs do not sensibly vary the relative weight of spinning with respect to tumbling and apparently approximately follow the simplistic  K41 prediction, eq (\ref{eq:k41_ratio}), and (red line) \ref{fig:VA_rotation_ratio}d. 
	
	\begin{figure*}
		\begin{center}
			\includegraphics[width=1.0\columnwidth]{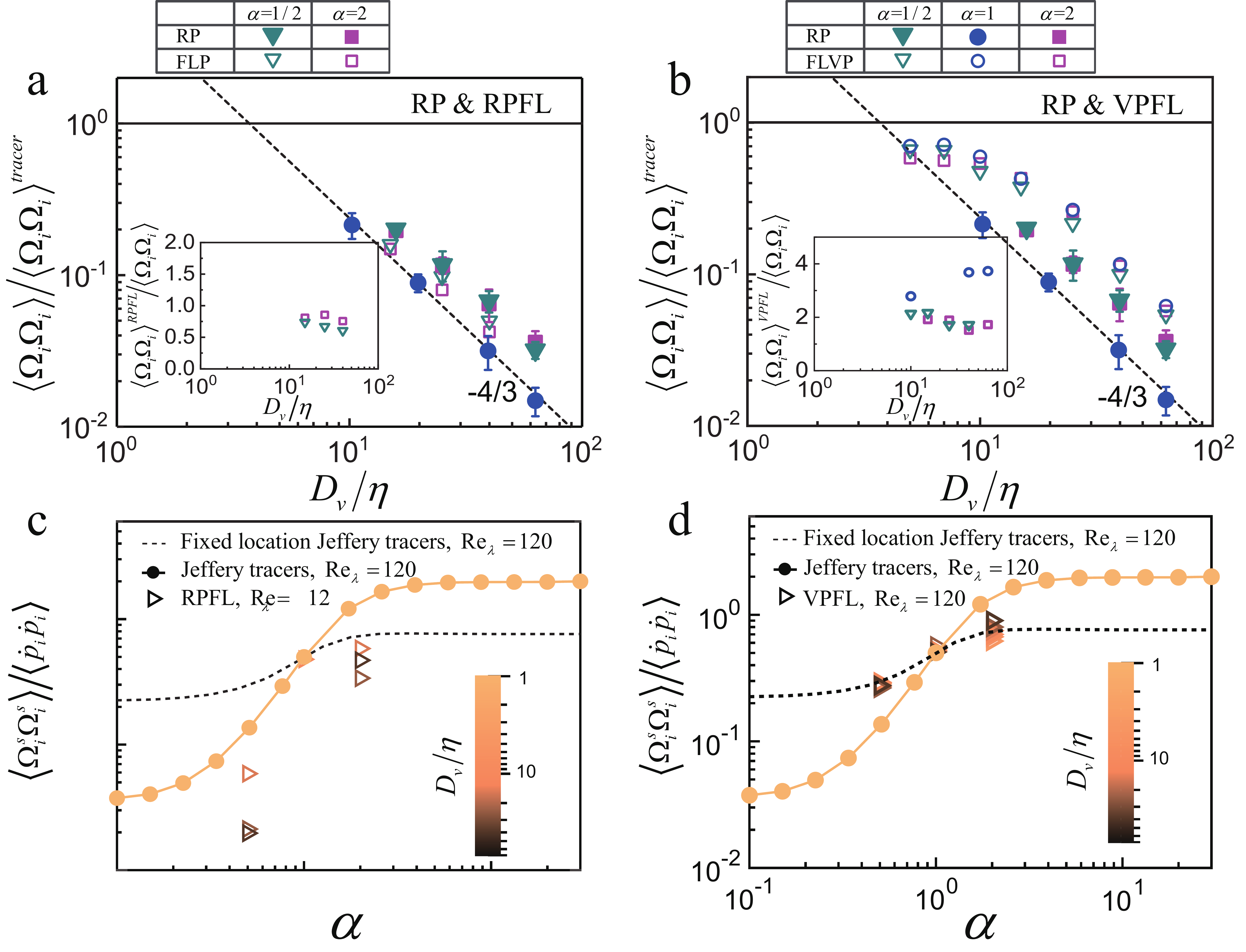}
			\caption{
			 Mean squared angular velocity, normalized by the results of tracers with corresponding aspect ratio, as a function of $D_v/\eta$ for the numerical experiments of (a) RPFL and (b) VPFL, compared with the results of RP. Insets show the ratio of the mean squared angular velocity of the numerical experiments to the results of RP as a function of $D_v/\eta$. The ratio of the mean spinning rate squared to the mean tumbling rate squared, $\langle \Omega_i^s\Omega_i^s\rangle /\langle \dot{p}_i\dot{p}_i\rangle$, as a function of aspect ratio $\alpha$ for the numerical experiments of (c) RPFL and (d) VPFL, respectively. The color bar represents the normalized particle size $D_v/\eta$ in the logarithmic scale. 
			}
			\label{fig:FixedPar}			
		\end{center}
	\end{figure*}
	
	In order to find out the leading physical mechanism responsible for the decrease of $\langle {\Omega}_i^s{\Omega}_i^s\rangle/\langle\dot{p}_i\dot{p}_i\rangle$, we carry out the numerical experiments with particles whose location is fixed, either with the solid-fluid interface or without it, denoted respectively as RPFL and VPFL, as shown in Figure~\ref{fig:FixedPar}. The rationale behind fixing the position is to rule out the role of preferential sampling of the flow position \citep{Eaton1991PreferentialConentration,Maxey1987PreferentialConcentration,calzavarini2008dimensionality}. Indeed, even if this has never been observed, for finite-sized spherical particles \cite{FiniteClusterPRE2012}, we can in principle not exclude it for non-spherical large particles. On the other hand, a direct measurement of preferential concentration in the present simulations is not an easy task due to the limited number of particles that we could place in the domain.
	
	We find that the mean normalized rotation rate variance of RPFL scales similarly RP and have a similar amplitude (see Figure~\ref{fig:FixedPar}a). Furthermore, $\langle {\Omega}_i^s{\Omega}_i^s\rangle/\langle\dot{p}_i\dot{p}_i\rangle$ decreases as the particle size increases (see Figure~\ref{fig:FixedPar}c), which is consistent with RP.
	These results indicate that the possible preferential sampling is not responsible for the decrease of $\langle {\Omega}_i^s{\Omega}_i^s\rangle/\langle\dot{p}_i\dot{p}_i\rangle$. 
	To further confirm this point, we show the rotational dynamics of VPFL in Figure~\ref{fig:FixedPar}(b and d). Similar scaling behavior of mean rotation rate square as a function of particle size is observed. And $\langle {\Omega}_i^s{\Omega}_i^s\rangle/\langle\dot{p}_i\dot{p}_i\rangle$ remains constant at the value of the fixed tracers for the corresponding aspect ratio. Since VP 
	do not have a solid interface and do not affect the flow, the results of these two additional numerical experiments suggest that the two-way coupling between the particles and the flow is responsible for the decrease of the ratio $\langle {\Omega}_i^s{\Omega}_i^s\rangle/\langle\dot{p}_i\dot{p}_i\rangle$. 
	
	\subsubsection{\ec{The role of particle boundary layers}}
    At this point, given the fact that fully resolved simulations allow virtually to measure any fluid quantity, one may wonder if it is possible to better characterize the effect of the particle feedback on the flow.
   With this in mind, we measure the fluid vorticity $\bm \omega$ and the rate of strain tensor $\mathcal{S}$ in the boundary layers surrounding the particles. We then compare the relative importance of their intensities. This is performed by computing the variance ratio $\tfrac{1}{4}\overline{\bm \omega^2} / \overline{\mathcal{S}:\mathcal{S}}$, where the average is performed over time, ensemble i.e. multiple trajectories, and over the azimuthal direction in order to improve the statistical convergence. We note that the ratio must have the value 1 in the particle-free HIT flow.

\begin{figure}
		\begin{center}
			\includegraphics[width=1.0\columnwidth]{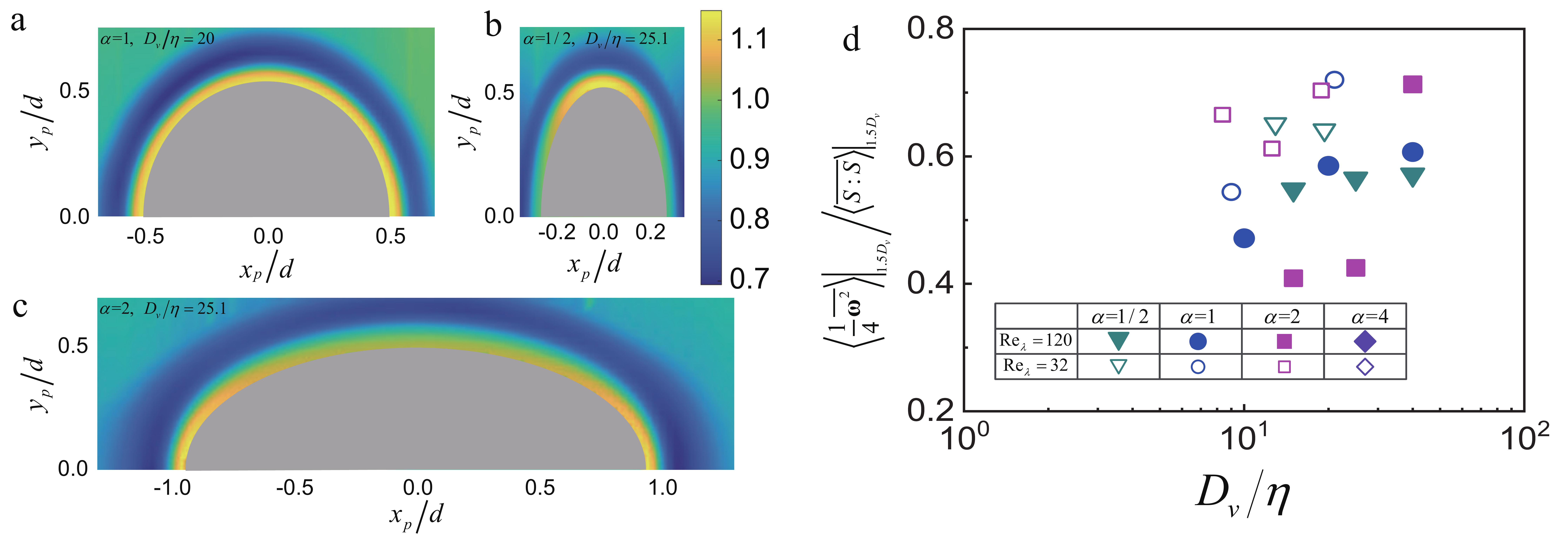}
			\caption{Field of the ratio $ \tfrac{1}{4} \overline{\bm \omega^2} /  \overline{\mathcal{S}:\mathcal{S}} $ at (a) $\alpha=1$, $D_v = 20 $ (b) $\alpha=0.5$, $D_v = 25.1$, and (c) $\alpha=2$, $D_v =25.1 $ at $Re_{\lambda} =120$. The colour bar indicates the value of $\tfrac{1}{4} \overline{\bm \omega^2} /  \overline{\mathcal{S}:\mathcal{S}}$. (d) The ratio of $\langle \tfrac{1}{4} \overline{\bm \omega^2} \rangle|_{1.5D_v} $ to $\langle \overline{\mathcal{S}:\mathcal{S}} \rangle|_{1.5D_v}$ as a function of $D_v/\eta$. 
			}
			\label{fig:W2_S2}			
		\end{center}
	\end{figure}

	Figure~\ref{fig:W2_S2}(a-c) shows three typical fields of the considered ratio around the particle up to the size of $1.5D_v$
	It is found that $\tfrac{1}{4} \overline{\bm \omega^2} /  \overline{\mathcal{S}:\mathcal{S}} \simeq 1.1$ (vorticity dominated) in a thin shell around the particle, while at immediately larger distances a wider region displays $\tfrac{1}{4} \overline{\bm \omega^2} / \ \overline{\mathcal{S}:\mathcal{S}} <1 $ (strain dominated), finally beyond $\sim 1.5 D_v$ any effect has vanished. 
	
	The quantity $\langle \tfrac{1}{4} \overline{\bm \omega^2} \rangle|_{1.5D_v} / \langle \overline{\mathcal{S}:\mathcal{S}} \rangle|_{1.5D_v}$, where $\langle \ldots \rangle_{1.5D_v}$ is a volume average over shells up to $1.5 D_v$,
	for different aspect ratios and particle sizes is reported in Figure~\ref{fig:W2_S2}(d).
	We can observe that the overall effect in the region affected by the particle presence is to decrease the importance of vorticity with respect to strain. 
	This points to the fact that spinning is possibly reduced as a consequence of the reduction of vorticity in the flow surrounding  a particle, in qualitative agreement with previous observations. However, we think that more sensitive tests remain to be performed in order to soundly quantify the effect of feedback on the flow. 
	
	\section{Conclusions}
	In summary, some statistical properties of the translational and rotational dynamics of neutrally buoyant finite-size spheroids in HIT have been investigated  by means of interface-resolved direct numerical simulations. Besides, the simulation of the real system, several extra numerical experiments have been carried out in order to enhance our insight on the essential features of the coupling of the particles with the flow.
	It is found that non-spherical inertial-scale particles experience spatially volume filtered turbulent fluctuations, similarly to spherical particles, even if the presence of finite-size particle sensibly influences the flow around the particle over a region that has similar aspect-ratio as the particle and approximately doubled $D_v$. 
	The variance of the acceleration of spheroids with different shapes collapses on a single line when plotted as a function of the equivalent diameter $D_v$, and at $Re_{\lambda}=120$ a power-law scaling $~D_v^{-1}$ is observed. It turns out that the acceleration variance versus particle size of virtual particles and fluid volume averages shows a similar scaling behavior, which proves that the scaling behavior of the acceleration intensity is dominated by the volume averaging of the acceleration of the flow field. Furthermore, the orientation of particles with size in the inertial range shows preferential alignments with flow structures. This is a novel result in the context of large particles, because it has been previously observed only for particles in the dissipative scale range of turbulence. We find that the preferential alignment of the particle acceleration shows a negligible size dependence, which suggests a similarity of flow coherent structures from the dissipative to inertial scales.

	Next, we studied the rotational dynamics of the particles as their size is increased. The mean square angular velocity of spheroids with different shapes show a clear scaling behavior with an exponent $~D_v^{-4/3}$, which agrees with predictions based on the K41 theory and with the available experiments \citep{ParsaPRL2014,bordoloi_variano_2017,OehmkePRF2021}. However, it turns out that the ratio of the mean square spinning rate to the mean square tumbling rate shows a systematic decrease as the particle size increases, and this is for all considered aspect ratios. To our knowledge, this new feature has never been observed before, and it calls for future experimental verifications.
	On the basis of the analysis of virtual particles, we interpret this phenomenon as due to the particle feedback on the flow. That is to say to the formation of boundary layers around the particles. In such layers the fluid strain-rate is overall increased in comparison to the vorticity, and this favors the tumbling with respect to the spinning. 
	
	The present findings improve the understanding of translational and rotational dynamics for neutrally-buoyant finite-size spheroids in turbulence. The results also shed light on a connection between the flow structures at the dissipative scales and inertial scales, which might open a new perspective for the investigation of turbulent flow. 
	We expect that a similar approach as the one followed in this study will also reaveal useful understanding of the dynamics of particles heavier/lighter than the carrying turbulent fluid flow.

	\section*{Acknowledgements}
	This work was supported by Natural Science Foundation of China under grant nos. 11988102 and 91852202, and Tencent Foundation through the {\sc xplorer prize}.

	\appendix
		\section{Particle acceleration statistics at $Re_{\lambda} = 32$}\label{sec:validation}
\begin{figure}
		\begin{center}
			\includegraphics[width=0.95\columnwidth]{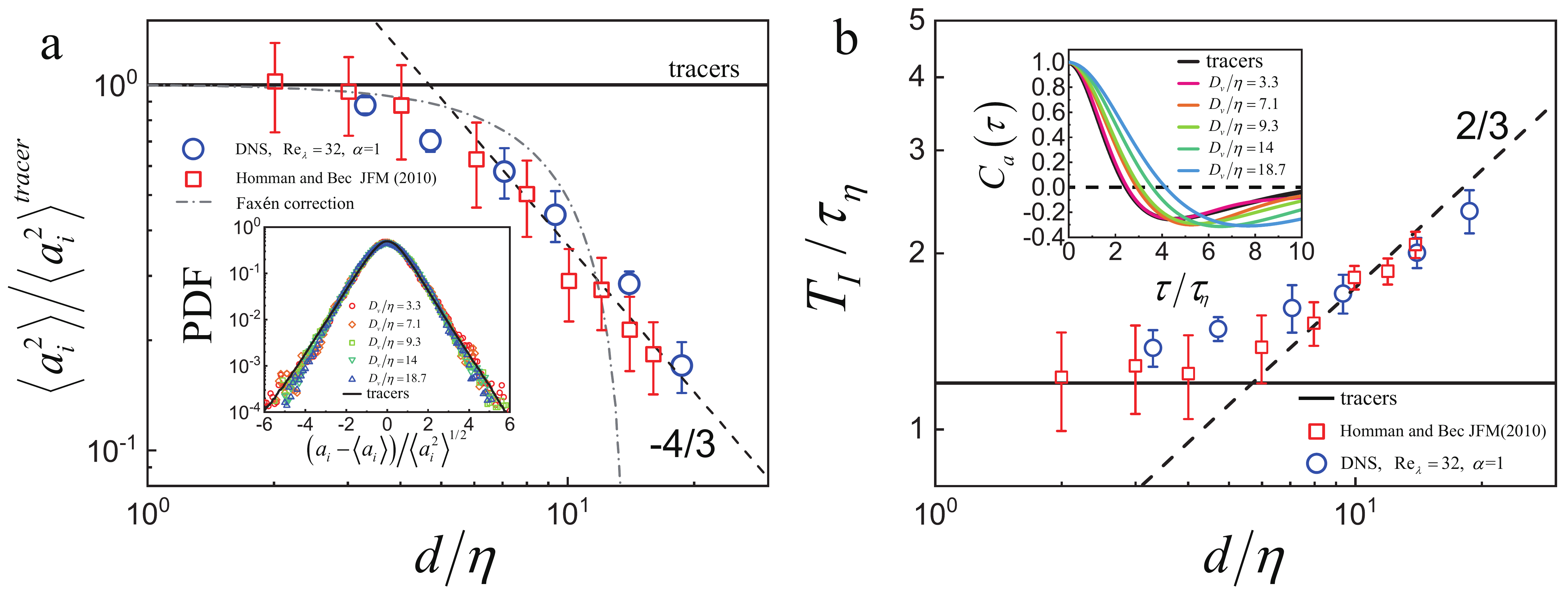}
			\caption{a). Normalized  single cartesian component of the acceleration variance $\langle a_i^2\rangle/\langle a_i^2\rangle^{tracer}$ as a function of the normalized particle size $d/\eta$ at $Re_\lambda=32$. Squares denote the results in Ref.~\citep{homann_bec_2010} at the same $Re_\lambda$. The dashed-dot line denotes the Fax\'en correction \citep{calzavarini2012impact}. The inset shows the PDFs of acceleration at different sizes. b). Normalized correlation time of acceleration, $T_I/\tau_\eta$, as a function of the normalized particle size $D_v/\eta$. Squares denote the correlation time in Ref.~\citep{homann_bec_2010} at the same $Re_\lambda$. The inset shows the autocorrelation function of a single cartesian acceleration component.
			}
			\label{fig:sm_validation}			
		\end{center}
	\end{figure}

	In order to validate the numerical results, we carried out simulations at $Re_\lambda=32$ for spheres, in the same condition as in a previous study \citep{homann_bec_2010}. Figure~\ref{fig:sm_validation}a shows the single-cartesian-component acceleration variance, $\langle a_i^2\rangle$, normalized by the acceleration variance of tracers, $\langle a_i^2\rangle^{tracer}$, as a function of normalized particle diameter $d/\eta$ ($d$ is the diameter of spheres). Our simulation results show good agreement with the previous study\citep{homann_bec_2010} and scale as $~(d/\eta)^{-4/3}$. At low $Re_\lambda$, the scaling exponent is smaller than -2/3 which is the exponent value expected in HIT with high $Re_\lambda$. The dashed-dot line represents the prediction of the Fax\'en correction based on point-particle simulation under the same $Re_\lambda$. When the particle is small ($d/\eta<4$), the acceleration variance is captured by the Fax\'en corrections. However, the Fax\'en correction can not account for the scaling behavior acceleration variance when $d/\eta>7$. The inset shows the PDFs of the acceleration component for the variant size of spheres. The effect of particle size on the PDF of acceleration is not significant, which is consistent with previous studies\citep{Xu2007prl,Brown2009prl,homann_bec_2010}. 
	Figure~\ref{fig:sm_validation}b shows the normalized integral time of 
	the temporal correlation function for particle acceleration, $T_I/\tau_{\eta}$, as a function $d/\eta$. The integral time $T_I$ is defined as the integral of the autocorrelation function from time zero to its first zero-crossing time $T_0$
	\begin{equation}
		T_{I}\equiv\int_{0}^{T_0}C(\tau)d\tau,~~ C(\tau)\equiv\frac{\langle a_i(t+\tau)a_i(t)\rangle}{\langle(a_i(t))^2\rangle}
		\label{eq:autoco}
	\end{equation}
	Our results are consistent with the correlation time in Ref.~\citep{homann_bec_2010} and show the power-law behaviour $T_I\sim(d)^{2/3}$. The inset shows the autocorrelation functions for various sizes of spheres, illustrating the increasing autocorrelation time as the size of the sphere increases. In conclusion, our translational statistics of simulation show good agreement with the previous study\citep{homann_bec_2010}, which validates our simulations.
\section{\ec{Comparison with \textit{Oehmke et al.}(2021) experiments}}\label{sec:comparison}	
\begin{figure}
		\begin{center}
			\includegraphics[width=1.0\columnwidth,angle=0]{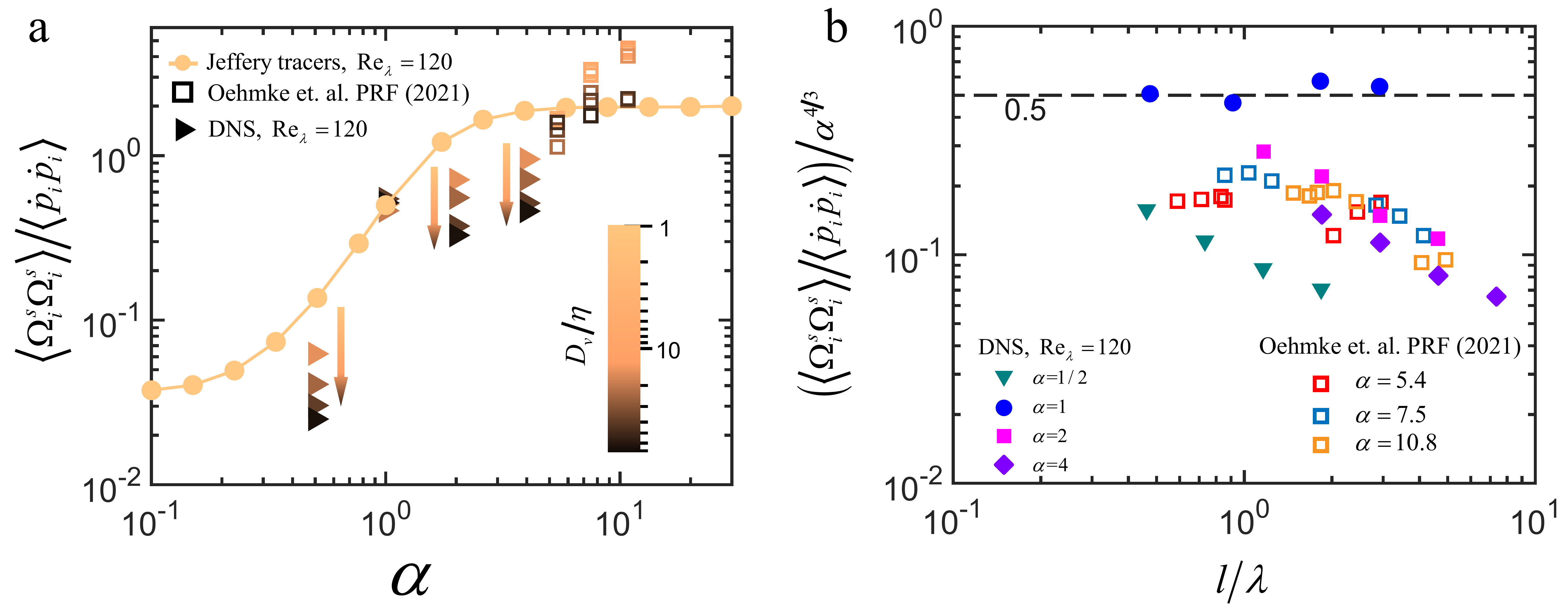}	
			\caption{\ec{Ratio of spinning rate variance over tumbling rate variance: comparison of present DNS results at $Re_{\lambda}=120$ with the experimental measurements by \cite{OehmkePRF2021} at $Re_{\lambda} \in \left[90,630\right]$. (a) Particle rotation ratio as a function of aspect ratio, here the color code maps the particle size in equivalent radius units;(b) Rotations ratio compensated by $\alpha^{4/3}$ as a function of the particle length (i.e. the symmetry axis) $l$ normalized by the Taylor microscale $\lambda$ (same as figure 5(b) in Oehmke et al. ~PRF 2021).}
			}
			\label{fig:ratio_cmp}			
		\end{center}
\end{figure}

\ec{We provide here a side-by-side comparison of the present simulations with the experiments by \cite{OehmkePRF2021}, which are the only to date to have measured both spinning and tumbling rates of particles. The experimental conditions of developed turbulence at $Re_{\lambda} \in \left[90,630\right]$ and particle aspect ratios $\alpha=5.4,7.5,10.8$ are relatively close to the ones explored in the present study. Figure \ref{fig:ratio_cmp}(a) shows the ratio of spinning rate variance over tumbling rate variance as a function of the aspect ratio. There is agreement on the dominance of spinning compared to tumbling for prolate ($\alpha >1$) particles (note that the isotropic value for spheres is 0.5). However, the decreasing trend at increasing the particle size $D_v$ which is clearly observed in the simulations is only partially consistent with the experiments. Figure \ref{fig:ratio_cmp}(b) represents the same measurements in a compensated form: the ratio of rotation rate variances is normalized by the $\alpha^{4/3}$ scaling as in the K41-based prediction Eq. (\ref{eq:k41_ratio}) and plotted against the length of the particle symmetry axis $l$ in $\lambda$ units. This choice of representation is adapted from figure 5(b) in \cite{OehmkePRF2021}. We observe here an excellent agreement for the behaviour of all prolate particles. However, the decreasing behaviour observed both for prolate and oblate particles stresses further the failure of the (\ref{eq:k41_ratio}) prediction.}  

	%\bibliographystyle{jfm}
	%\bibliography{finite-spheroid}

\begin{thebibliography}{55}
\expandafter\ifx\csname natexlab\endcsname\relax\def\natexlab#1{#1}\fi
\def\au#1{#1} \def\ed#1{#1} \def\yr#1{#1}\def\at#1{#1}\def\jt#1{\textit{#1}}
  \def\bt#1{#1}\def\bvol#1{\textbf{#1}} \def\vol#1{#1} \def\pg#1{#1}
  \def\publ#1{#1}\def\arxiv#1{#1}\def\org#1{#1}\def\st#1{\textit{#1}}

\bibitem[Ardekani {\em et~al.\/}(2016)Ardekani, Costa, Breugem \&
  Brandt]{luca2016IJMF}
{\sc \au{Ardekani, M.~N.}, \au{Costa, P.}, \au{Breugem, W.~P.} \& \au{Brandt,
  L.}} \yr{2016}  \at{Numerical study of the sedimentation of spheroidal
  particles}.  \jt{Int. J. Multiph. Flow}  \bvol{87},  \pg{16--34}.

\bibitem[Bec {\em et~al.\/}(2007)Bec, Biferale, Cencini, Lanotte, Musacchio \&
  Toschi]{bec2007heavy}
{\sc \au{Bec, J.}, \au{Biferale, L.}, \au{Cencini, M.}, \au{Lanotte, A.},
  \au{Musacchio, S.} \& \au{Toschi, F.}} \yr{2007}  \at{Heavy particle
  concentration in turbulence at dissipative and inertial scales}.  \jt{Phys.
  Rev. Lett.}  \bvol{98}~(8),  \pg{084502}.

\bibitem[Bentkamp {\em et~al.\/}(2019)Bentkamp, Lalescu \&
  Wilczek]{bentkamp2019persistent}
{\sc \au{Bentkamp, L.}, \au{Lalescu, C.~C.} \& \au{Wilczek, M.}} \yr{2019}
  \at{Persistent accelerations disentangle lagrangian turbulence}.  \jt{Nat.
  Commun.}  \bvol{10}~(1),  \pg{1--8}.

\bibitem[Benzi {\em et~al.\/}(2009)Benzi, Biferale, Calzavarini, Lohse \&
  Toschi]{BenziPRE2009}
{\sc \au{Benzi, R.}, \au{Biferale, L.}, \au{Calzavarini, E.}, \au{Lohse, D.} \&
  \au{Toschi, F.}} \yr{2009}  \at{Velocity-gradient statistics along particle
  trajectories in turbulent flows: The refined similarity hypothesis in the
  lagrangian frame}.  \jt{Phys. Rev. E}  \bvol{80},  \pg{066318}.

\bibitem[Biferale {\em et~al.\/}(2005)Biferale, Boffetta, Celani, Lanotte \&
  Toschi]{biferale2005trap}
{\sc \au{Biferale, L.}, \au{Boffetta, G.}, \au{Celani, A.}, \au{Lanotte, A.} \&
  \au{Toschi, F.}} \yr{2005}  \at{Particle trapping in three-dimensional fully
  developed turbulence}.  \jt{Phys. Fluids}  \bvol{17}~(2),  \pg{021701}.

\bibitem[Bordoloi \& Variano(2017)]{bordoloi_variano_2017}
{\sc \au{Bordoloi, A.~D.} \& \au{Variano, E.}} \yr{2017}  \at{Rotational
  kinematics of large cylindrical particles in turbulence}.  \jt{J. Fluid
  Mech.}  \bvol{815},  \pg{199--222}.

\bibitem[Bounoua {\em et~al.\/}(2018)Bounoua, Bouchet \&
  Verhille]{BounouaPRL2018}
{\sc \au{Bounoua, S.}, \au{Bouchet, G.} \& \au{Verhille, G.}} \yr{2018}
  \at{Tumbling of inertial fibers in turbulence}.  \jt{Phys. Rev. Lett.}
  \bvol{121},  \pg{124502}.

\bibitem[Brenner(1961)]{Brenner1961}
{\sc \au{Brenner, H.}} \yr{1961}  \at{The slow motion of a sphere through a
  viscous fluid towards a plane surface}.  \jt{Chem. Eng. Sci.}  \bvol{16}~(3),
   \pg{242--251}.

\bibitem[Brown {\em et~al.\/}(2009)Brown, Warhaft \& Voth]{Brown2009prl}
{\sc \au{Brown, R.~D.}, \au{Warhaft, Z.} \& \au{Voth, G.~A.}} \yr{2009}
  \at{Acceleration statistics of neutrally buoyant spherical particles in
  intense turbulence}.  \jt{Phys. Rev. Lett.}  \bvol{103},  \pg{194501}.

\bibitem[{Brändle de Motta} {\em et~al.\/}(2016){Brändle de Motta},
  Estivalezes, Climent \& Vincent]{BRANDLEDEMOTTA2016369}
{\sc \au{{Brändle de Motta}, J.C.}, \au{Estivalezes, J.L.}, \au{Climent, E.}
  \& \au{Vincent, S.}} \yr{2016}  \at{Local dissipation properties and
  collision dynamics in a sustained homogeneous turbulent suspension composed
  of finite size particles}.  \jt{International Journal of Multiphase Flow}
  \bvol{85},  \pg{369--379}.

\bibitem[Byron {\em et~al.\/}(2015)Byron, Einarsson, Gustavsson, Voth, Mehlig
  \& Variano]{Byron2015}
{\sc \au{Byron, M.}, \au{Einarsson, J.}, \au{Gustavsson, K.}, \au{Voth, G.~A.},
  \au{Mehlig, B.} \& \au{Variano, E.}} \yr{2015}  \at{Shape-dependence of
  particle rotation in isotropic turbulence}.  \jt{Phys. Fluids}
  \bvol{27}~(3),  \pg{035101}.

\bibitem[Calzavarini(2019)]{Calzavarini_SI2019}
{\sc \au{Calzavarini, E.}} \yr{2019}  \at{Eulerian-lagrangian fluid dynamics
  platform: The ch4-project}.  \jt{Software Impacts}  \bvol{1},  \pg{100002}.

\bibitem[Calzavarini {\em et~al.\/}(2020)Calzavarini, Jiang \&
  Sun]{calzavarini2020anisotropic}
{\sc \au{Calzavarini, E.}, \au{Jiang, L.} \& \au{Sun, C.}} \yr{2020}
  \at{Anisotropic particles in two-dimensional convective turbulence}.
  \jt{Phys. Fluids}  \bvol{32}~(2),  \pg{023305}.

\bibitem[Calzavarini {\em et~al.\/}(2008)Calzavarini, Kerscher, Lohse \&
  Toschi]{calzavarini2008dimensionality}
{\sc \au{Calzavarini, E.}, \au{Kerscher, M.}, \au{Lohse, D.} \& \au{Toschi,
  F.}} \yr{2008}  \at{Dimensionality and morphology of particle and bubble
  clusters in turbulent flow}.  \jt{J. Fluid Mech.}  \bvol{607},  \pg{13--24}.

\bibitem[Calzavarini {\em et~al.\/}(2009)Calzavarini, Volk, Bourgoin, Leveque,
  Pinton \& Toschi]{calzavarini2009acceleration}
{\sc \au{Calzavarini, E.}, \au{Volk, R.}, \au{Bourgoin, M.}, \au{Leveque, E.},
  \au{Pinton, J.~F.} \& \au{Toschi, F.}} \yr{2009}  \at{Acceleration statistics
  of finite-sized particles in turbulent flow: the role of fax{\'e}n forces}.
  \jt{J. Fluid Mech.}  \bvol{630},  \pg{179--189}.

\bibitem[Calzavarini {\em et~al.\/}(2012)Calzavarini, Volk, Leveque, Pinton \&
  Toschi]{calzavarini2012impact}
{\sc \au{Calzavarini, E.}, \au{Volk, R.}, \au{Leveque, E.}, \au{Pinton, J.~F.}
  \& \au{Toschi, F.}} \yr{2012}  \at{Impact of trailing wake drag on the
  statistical properties and dynamics of finite-sized particle in turbulence}.
  \jt{Phys. D}  \bvol{241D}~(3),  \pg{237--244}.

\bibitem[Chong {\em et~al.\/}(1990)Chong, Perry \& Cantwell]{Chong1990A}
{\sc \au{Chong, M.~S.}, \au{Perry, A.~E.} \& \au{Cantwell, B.~J.}} \yr{1990}
  \at{A general classification of three‐dimensional flow fields}.  \jt{Phys.
  Fluids}  \bvol{2}~(5),  \pg{765--777}.

\bibitem[Cisse {\em et~al.\/}(2013)Cisse, Homann \& Bec]{cisse2013slipping}
{\sc \au{Cisse, M.}, \au{Homann, H.} \& \au{Bec, J.}} \yr{2013}  \at{Slipping
  motion of large neutrally buoyant particles in turbulence}.  \jt{J. Fluid
  Mech.}  \bvol{735},  \pg{R1}.

\bibitem[Cooley \& O'Neill(1969)]{cooley1969}
{\sc \au{Cooley, M. D.~A.} \& \au{O'Neill, M.~E.}} \yr{1969}  \at{On the slow
  motion generated in a viscous fluid by the approach of a sphere to a plane
  wall or stationary sphere}.  \jt{Mathematika}  \bvol{16}~(1),  \pg{37--49}.

\bibitem[Costa {\em et~al.\/}(2015)Costa, Boersma, Westerweel \&
  Breugem]{PRE2015collision}
{\sc \au{Costa, P.}, \au{Boersma, B.~J.}, \au{Westerweel, J.} \& \au{Breugem,
  W.~P.}} \yr{2015}  \at{Collision model for fully resolved simulations of
  flows laden with finite-size particles}.  \jt{Phys. Rev. E}  \bvol{92},
  \pg{053012}.

\bibitem[Do-Quang {\em et~al.\/}(2014)Do-Quang, Amberg, Brethouwer \&
  Johansson]{Quang2014PRE}
{\sc \au{Do-Quang, M.}, \au{Amberg, G.}, \au{Brethouwer, G.} \& \au{Johansson,
  A.~V.}} \yr{2014}  \at{Simulation of finite-size fibers in turbulent channel
  flows}.  \jt{Phys. Rev. E}  \bvol{89},  \pg{013006}.

\bibitem[Dolata \& Zia(2021)]{dolata_zia_2021}
{\sc \au{Dolata, B.~E.} \& \au{Zia, R.~N.}} \yr{2021}  \at{Fax\'en formulas for
  particles of arbitrary shape and material composition}.  \jt{Journal of Fluid
  Mechanics}  \bvol{910},  \pg{A22}.

\bibitem[Fiabane {\em et~al.\/}(2012)Fiabane, Zimmermann, Volk, Pinton \&
  Bourgoin]{FiniteClusterPRE2012}
{\sc \au{Fiabane, L.}, \au{Zimmermann, R.}, \au{Volk, R.}, \au{Pinton, J.~F.}
  \& \au{Bourgoin, M.}} \yr{2012}  \at{Clustering of finite-size particles in
  turbulence}.  \jt{Phys. Rev. E}  \bvol{86},  \pg{035301}.

\bibitem[Gustavsson {\em et~al.\/}(2014)Gustavsson, Einarsson \&
  Mehlig]{Gustavsson2014}
{\sc \au{Gustavsson, K.}, \au{Einarsson, J.} \& \au{Mehlig, B.}} \yr{2014}
  \at{Tumbling of small axisymmetric particles in random and turbulent flows}.
  \jt{Phys. Rev. Lett.}  \bvol{112},  \pg{014501}.

\bibitem[Homann \& Bec(2010)]{homann_bec_2010}
{\sc \au{Homann, H.} \& \au{Bec, J.}} \yr{2010}  \at{Finite-size effects in the
  dynamics of neutrally buoyant particles in turbulent flow}.  \jt{J. Fluid
  Mech.}  \bvol{651},  \pg{81--91}.

\bibitem[Jeffery(1922)]{Jeffery1922}
{\sc \au{Jeffery, G.~B.}} \yr{1922}  \at{The motion of ellipsoidal particles
  immersed in a viscous fluid}.  \jt{Proc. R. Soc. Lond. A}  \bvol{102}~(715),
  \pg{161--179}.

\bibitem[Jiang {\em et~al.\/}(2020)Jiang, Calzavarini \&
  Sun]{jiang2020rotation}
{\sc \au{Jiang, L.}, \au{Calzavarini, E.} \& \au{Sun, C.}} \yr{2020}
  \at{Rotation of anisotropic particles in rayleigh--b{\'e}nard turbulence}.
  \jt{J. Fluid Mech.}  \bvol{901},  \pg{A8}.

\bibitem[Jiang {\em et~al.\/}(2021)Jiang, Wang, Liu \&
  Calzavarini]{jiang2021rotational}
{\sc \au{Jiang, L.}, \au{Wang, C.}, \au{Liu, S.and~Sun, C.} \& \au{Calzavarini,
  E.}} \yr{2021}  \at{Rotational dynamics of bottom-heavy rods in turbulence
  from experiments and numerical simulations}.  \jt{Theor. App. Mech. Lett.}
  \bvol{11},  \pg{100227}.

\bibitem[Kuperman {\em et~al.\/}(2019)Kuperman, Sabban \& van
  Hout]{Kuperman2019prf}
{\sc \au{Kuperman, S.}, \au{Sabban, L.} \& \au{van Hout, R.}} \yr{2019}
  \at{Inertial effects on the dynamics of rigid heavy fibers in isotropic
  turbulence}.  \jt{Phys. Rev. Fluids}  \bvol{4},  \pg{064301}.

\bibitem[Liberzon {\em et~al.\/}(2012)Liberzon, Luthi, Holzner, Ott, Berg \&
  Jakob]{Liberzon_acceleration2012}
{\sc \au{Liberzon, A.}, \au{Luthi, B.}, \au{Holzner, M.}, \au{Ott, S.},
  \au{Berg, J.} \& \au{Jakob, M.}} \yr{2012}  \at{On the structure of
  acceleration in turbulence}.  \jt{Phys. D}  \bvol{241}~(3),  \pg{208--215}.

\bibitem[Luo {\em et~al.\/}(2007)Luo, Wang, Fan \& Cen]{PRE_multiforce2007}
{\sc \au{Luo, K.}, \au{Wang, Z.}, \au{Fan, J.} \& \au{Cen, K.}} \yr{2007}
  \at{Full-scale solutions to particle-laden flows: Multidirect forcing and
  immersed boundary method}.  \jt{Phys. Rev. E}  \bvol{76},  \pg{066709}.

\bibitem[Mathai {\em et~al.\/}(2016)Mathai, Calzavarini, Brons, Sun \&
  Lohse]{mathai2016microbubbles}
{\sc \au{Mathai, V.}, \au{Calzavarini, E.}, \au{Brons, J.}, \au{Sun, C.} \&
  \au{Lohse, D.}} \yr{2016}  \at{Microbubbles and microparticles are not
  faithful tracers of turbulent acceleration}.  \jt{Phys. Rev. Lett.}
  \bvol{117}~(2),  \pg{024501}.

\bibitem[Mathai {\em et~al.\/}(2018)Mathai, Huisman, Sun, Lohse \&
  Bourgoin]{mathai2018dispersion}
{\sc \au{Mathai, V.}, \au{Huisman, S.~G.}, \au{Sun, C.}, \au{Lohse, D.} \&
  \au{Bourgoin, M.}} \yr{2018}  \at{Dispersion of air bubbles in isotropic
  turbulence}.  \jt{Phys. Rev. Lett.}  \bvol{121}~(5),  \pg{054501}.

\bibitem[Maxey(1987)]{Maxey1987PreferentialConcentration}
{\sc \au{Maxey, M.~R.}} \yr{1987}  \at{The gravitational settling of aerosol
  particles in homogeneous turbulence and random flow fields}.  \jt{J. Fluid
  Mech.}  \bvol{174},  \pg{441--465}.

\bibitem[Mittal \& Iaccarino(2005)]{mittal2005immersed}
{\sc \au{Mittal, R.} \& \au{Iaccarino, G.}} \yr{2005}  \at{Immersed boundary
  methods}.  \jt{Annu. Rev. Fluid Mech.}  \bvol{37},  \pg{239--261}.

\bibitem[Ni {\em et~al.\/}(2014)Ni, Ouellette \& Voth]{ni_ouellette_voth_2014}
{\sc \au{Ni, Rui}, \au{Ouellette, Nicholas~T.} \& \au{Voth, Greg~A.}} \yr{2014}
   \at{Alignment of vorticity and rods with lagrangian fluid stretching in
  turbulence}.  \jt{J. Fluid Mech.}  \bvol{743},  \pg{R3}.

\bibitem[Oehmke {\em et~al.\/}(2021)Oehmke, Bordoloi, Variano \&
  Verhille]{OehmkePRF2021}
{\sc \au{Oehmke, T.~B.}, \au{Bordoloi, A.~D.}, \au{Variano, E.} \&
  \au{Verhille, G.}} \yr{2021}  \at{Spinning and tumbling of long fibers in
  isotropic turbulence}.  \jt{Phys. Rev. Fluids}  \bvol{6},  \pg{044610}.

\bibitem[Parsa {\em et~al.\/}(2012)Parsa, Calzavarini, Toschi \&
  Voth]{ParsaPRL2012}
{\sc \au{Parsa, S.}, \au{Calzavarini, E.}, \au{Toschi, F.} \& \au{Voth, G.~A.}}
  \yr{2012}  \at{Rotation rate of rods in turbulent fluid flow}.  \jt{Phys.
  Rev. Lett.}  \bvol{109}~(13),  \pg{134501}.

\bibitem[Parsa \& Voth(2014)]{ParsaPRL2014}
{\sc \au{Parsa, S.} \& \au{Voth, G.~A.}} \yr{2014}  \at{Inertial range scaling
  in rotations of long rods in turbulence}.  \jt{Phys. Rev. Lett.}  \bvol{112},
   \pg{024501}.

\bibitem[Peskin(2002)]{peskin_2002}
{\sc \au{Peskin, C.~S.}} \yr{2002}  \at{The immersed boundary method}.
  \jt{Acta Numer.}  \bvol{11},  \pg{479--517}.

\bibitem[Porta {\em et~al.\/}(2001)Porta, Voth, Crawford, Alexander \&
  Bodenschatz]{2001NatureParticle}
{\sc \au{Porta, A.~La}, \au{Voth, G.~A.}, \au{Crawford, A.~M.}, \au{Alexander,
  J.} \& \au{Bodenschatz, E.}} \yr{2001}  \at{Fluid particle accelerations in
  fully developed turbulence}.  \jt{Nature}  \bvol{409},  \pg{22}.

\bibitem[Pujara {\em et~al.\/}(2018)Pujara, Oehmke, Bordoloi \&
  Variano]{Pujara2018prf}
{\sc \au{Pujara, N.}, \au{Oehmke, T.~B.}, \au{Bordoloi, A.~D.} \& \au{Variano,
  E.~A.}} \yr{2018}  \at{Rotations of large inertial cubes, cuboids, cones, and
  cylinders in turbulence}.  \jt{Phys. Rev. Fluids}  \bvol{3},  \pg{054605}.

\bibitem[Pujara {\em et~al.\/}(2019)Pujara, Voth \&
  Variano]{pujara_voth_variano_2019}
{\sc \au{Pujara, Nimish}, \au{Voth, Greg~A.} \& \au{Variano, Evan~A.}}
  \yr{2019}  \at{Scale-dependent alignment, tumbling and stretching of slender
  rods in isotropic turbulence}.  \jt{J. Fluid Mech.}  \bvol{860},
  \pg{465–486}.

\bibitem[Pumir \& Wilkinson(2011)]{PumirWilkinson2011}
{\sc \au{Pumir, A.} \& \au{Wilkinson, M.}} \yr{2011}  \at{Orientation
  statistics of small particles in turbulence}.  \jt{New J. Phys.}
  \bvol{13}~(9),  \pg{093030}.

\bibitem[Qureshi {\em et~al.\/}(2007)Qureshi, Bourgoin, Baudet, Cartellier \&
  Gagne]{qureshi2007turbulent}
{\sc \au{Qureshi, N.~M.}, \au{Bourgoin, M.}, \au{Baudet, C.}, \au{Cartellier,
  A.} \& \au{Gagne, Y.}} \yr{2007}  \at{Turbulent transport of material
  particles: an experimental study of finite size effects}.  \jt{Phys. Rev.
  Lett.}  \bvol{99}~(18),  \pg{184502}.

\bibitem[Shen {\em et~al.\/}(2021)Shen, Lu, Wang \& Peng]{WangPRE2021}
{\sc \au{Shen, J.}, \au{Lu, Z.}, \au{Wang, L.} \& \au{Peng, C.}} \yr{2021}
  \at{Influence of particle-fluid density ratio on the dynamics of finite-size
  particles in homogeneous isotropic turbulent flows}.  \jt{Phys. Rev. E}
  \bvol{104},  \pg{025109}.

\bibitem[Squires \& Eaton(1991)]{Eaton1991PreferentialConentration}
{\sc \au{Squires, K.~D.} \& \au{Eaton, J.~K.}} \yr{1991}  \at{Preferential
  concentration of particles by turbulence}.  \jt{Phys. Fluids}  \bvol{3},
  \pg{1169}.

\bibitem[Suzuki \& Inamuro(2011)]{SUZUKI2011173}
{\sc \au{Suzuki, K.} \& \au{Inamuro, T.}} \yr{2011}  \at{Effect of internal
  mass in the simulation of a moving body by the immersed boundary method}.
  \jt{Comput. Fluids}  \bvol{49}~(1),  \pg{173--187}.

\bibitem[Uhlmann(2005)]{uhlmann2005}
{\sc \au{Uhlmann, M.}} \yr{2005}  \at{An immersed boundary method with direct
  forcing for the simulation of particulate flows}.  \jt{J. Comput. Phys.}
  \bvol{209}~(2),  \pg{448--476}.

\bibitem[Volk {\em et~al.\/}(2011)Volk, Calzavarini, Leveque \&
  Pinton]{volk2011dynamics}
{\sc \au{Volk, R.}, \au{Calzavarini, E.}, \au{Leveque, E.} \& \au{Pinton,
  J.~F.}} \yr{2011}  \at{Dynamics of inertial particles in a turbulent von
  K{\'a}rm{\'a}n flow}.  \jt{J. Fluid Mech.}  \bvol{668},  \pg{223--235}.

\bibitem[Voth {\em et~al.\/}(2002)Voth, Porta, Crawford, Alexander \&
  Bodenschatz]{voth_laporta_crawford_alexander_bodenschatz_2002}
{\sc \au{Voth, G.~A.}, \au{Porta, L.~A.}, \au{Crawford, A.~M.}, \au{Alexander,
  J.} \& \au{Bodenschatz, E.}} \yr{2002}  \at{Measurement of particle
  accelerations in fully developed turbulence}.  \jt{J. Fluid Mech.}
  \bvol{469},  \pg{121--160}.

\bibitem[Voth \& Soldati(2017)]{VothARFM2017}
{\sc \au{Voth, G.~A.} \& \au{Soldati, A.}} \yr{2017}  \at{Anisotropic particles
  in turbulence}.  \jt{Annu. Rev. Fluid Mech.}  \bvol{49},  \pg{249--276}.

\bibitem[Wang {\em et~al.\/}(2021)Wang, Jiang, Jiang, Sun \& Liu]{Cheng2021JHD}
{\sc \au{Wang, C.}, \au{Jiang, L.}, \au{Jiang, H.}, \au{Sun, C.} \& \au{Liu,
  S.}} \yr{2021}  \at{Heat transfer and flow structure of two-dimensional
  thermal convection overratchet surfaces}.  \jt{J. Hydrodyn.}  \pg{p. In
  Press}.

\bibitem[Xu \& Bodenschatz(2008)]{xu2008motion}
{\sc \au{Xu, H.} \& \au{Bodenschatz, E.}} \yr{2008}  \at{Motion of inertial
  particles with size larger than kolmogorov scale in turbulent flows}.
  \jt{Phys. D}  \bvol{237}~(14-17),  \pg{2095--2100}.

\bibitem[Xu {\em et~al.\/}(2007)Xu, Ouellette, Vincenzi \&
  Bodenschatz]{Xu2007prl}
{\sc \au{Xu, H.}, \au{Ouellette, N.~T.}, \au{Vincenzi, D.} \& \au{Bodenschatz,
  E.}} \yr{2007}  \at{Acceleration correlations and pressure structure
  functions in high-reynolds number turbulence}.  \jt{Phys. Rev. Lett.}
  \bvol{99},  \pg{204501}.

\end{thebibliography}

\end{document}